\documentclass[usenatbib,twocolumn]{mn2e}
\usepackage{times}
\usepackage{graphicx}
\usepackage{amsfonts}
\usepackage{epsfig,rotating,natbib,pstricks}
\usepackage{color}
\usepackage{amssymb,latexsym}
\usepackage{amsmath}

\pubyear{2008}
\def\N{$\mathrm N$}
 
\def\beq{\begin{equation}}
\def\eeq{\end{equation}} 
\def\baq{\begin{eqnarray}}
\def\eaq{\end{eqnarray}}

\def\bv{{\mathbf{v}}}

\def\br{{\mathbf{r}}}

\def\bF{{\mathbf{F}}}
\def\epsmax{\epsilon_{max}}
\def\epsmin{\epsilon_{min}}
\def\epsofr{\epsilon(\br)}
\def\gradeps{\nabla \epsilon}
\def\gradphi{\nabla \phi}
\def\p3m{P$^3$M}
\def\ap3m{AP$^3$M}

\def\xibar{\bar{\xi}}
\def\rnl{r_{nl}}
\def\rss{r_{ss}}
\def\rcon{r_{con}}
\def\barrij{\bar{r}_{ij}}
\def\cpricet{\citet{2007MNRAS.374.1347P}}

\def\dr{\!{\rm d}^3\!\mathbf{r}}
\def\dv{\!{\rm d}^3\!\mathbf{v}}

\newcommand{\pder}[2]{\ensuremath{\frac{\partial #1}{\partial #2}}}

\title[Adaptive TreePM]{The Adaptive TreePM: An Adaptive Resolution 
Code for Cosmological \N-body Simulations}

\author[Bagla and Khandai]%
{J. S. Bagla and Nishikanta Khandai \\
Harish-Chandra Research Institute, Chhatnag Road, Jhunsi, \\
Allahabad 211019, INDIA \\
E-mail: jasjeet@hri.res.in, nishi@hri.res.in}  
\pubyear{2008}

\label{firstpage}

\def\LaTeX{L\kern-.36em\raise.3ex\hbox{a}\kern-.15em
    T\kern-.1667em\lower.7ex\hbox{E}\kern-.125emX}

\pagerange{\pageref{firstpage}--\pageref{lastpage}} 

\begin{document}

\maketitle

\begin{abstract}
Cosmological \N-Body simulations are used for a variety of applications.
Indeed progress in the study of large scale structures and galaxy
formation would have been very limited without this tool. 
For nearly twenty years the limitations imposed by computing power
forced simulators to ignore some of the basic requirements for
modeling gravitational instability.
One of the limitations of most cosmological codes has been the use of
a force softening length that is much smaller than the typical
inter-particle separation.
This leads to departures from collisionless evolution that is desired
in these simulations.
We propose a particle based method with an adaptive resolution
where the force softening length is reduced in high density regions
while ensuring that it remains well above the local inter-particle
separation. 
The method, called the Adaptive TreePM, is based on the TreePM code. 
We present the mathematical model and an implementation of this code,
and demonstrate that the results converge over a range of options for
parameters introduced in generalizing the code from the TreePM code.
We explicitly demonstrate collisionless evolution in collapse of an oblique
plane wave. 
We compare the code with the fixed resolution TreePM code and also an
implementation that mimics adaptive mesh refinement methods and
comment on the agreement, and disagreements in the results. 
We find that in most respects the ATreePM code performs at least as well as
the fixed resolution TreePM in highly over-dense regions, from clustering and 
number density of haloes, to internal dynamics of haloes. 
We also show that the adaptive code is faster than the corresponding high
resolution TreePM code.
\end{abstract}




\section{Introduction}

Large scale structures traced by galaxies are believed to have formed by
amplification of small perturbations
\citep{1980lssu.book.....P,1999coph.book.....P,2002tagc.book.....P,2002PhR...367....1B}.     
Galaxies are highly over-dense systems, matter density $\rho$ in galaxies is
thousands of times larger than the average density $\bar\rho$ in the
universe.  
Typical density contrast ($\delta \equiv \rho/{\bar\rho} - 1$) in matter at
these scales in the early universe was much smaller than unity.  
Thus the problem of galaxy formation and the large scale distribution of
galaxies requires an understanding of the evolution of density perturbations
from small initial values to the large values we encounter today.

Initial density perturbations were present at all scales that have been
observed
\citep{2007ApJS..170..377S,2007ApJ...657..645P}.
The equations that describe the evolution of density perturbations in an
expanding universe have been known for several decades
\citep{1974A&A....32..391P} and these are easy to solve when the amplitude of
perturbations is small.   
Once density contrast at relevant scales becomes comparable to unity, 
perturbations becomes non-linear and coupling with perturbations at other
scales cannot be ignored.  
The equation for evolution of density perturbations cannot be solved for
generic initial conditions in this regime.
N-Body simulations (e.g., see
\citet{1985ApJS...57..241E,1998ARA&A..36..599B,1997Prama..49..161B,2005CSci...88.1088B,2008SSRv..134..229D})
are often used to study the evolution in this regime.  
Alternative approaches can be used if one requires only a limited
amount of information and in such a case either quasi-linear approximation
schemes 
\citep{2002PhR...367....1B,1970A&A.....5...84Z,1989MNRAS.236..385G,1992MNRAS.259..437M,1993ApJ...418..570B,1994MNRAS.266..227B,1995PhR...262....1S,1996ApJ...471....1H}
or scaling relations
\citep{1977ApJS...34..425D,1991ApJ...374L...1H,1995MNRAS.276L..25J,2000ApJ...531...17K,1998ApJ...508L...5M,1994MNRAS.271..976N,1996ApJ...466..604P,1994MNRAS.267.1020P,1996MNRAS.278L..29P,1996MNRAS.280L..19P,2003MNRAS.341.1311S}
suffice. 
However, even the approximation schemes and scaling relations must be compared
and calibrated with simulations before these can be used with confidence.

Last three decades have seen a rapid development of techniques and computing
power for cosmological simulations and the results of these simulations have
provided valuable insight into the study of structure formation.
The state of the art simulations used less than $10^5$ particles two
decades ago \citep{1988MNRAS.235..715E} and if the improvement
had been due only to computing power then the largest simulation possible
today should have been around $10^9$ particles, whereas the largest simulations
done till date used more than $10^{10}$ particles \citep{2005Natur.435..629S}. 
Evidently, development of new methods and optimizations has also played a
significant role in the evolution of simulation studies
\citep{1985ApJS...57..241E,1986Natur.324..446B,1987JCoPh..73..325G,1988ApJS...68..521B,1989ApJS...71..871J,1990JCoPh..87..137H,1990JCoPh..87..148M,1991PASJ...43..621M,1991ApJS...75..231H,1991ApJ...368L..23C,1993PASJ...45..269E,1994CoPhC..78..238T,1995ApJ...453..566B,1995MNRAS.274..287S,1995ApJS...98..355X,1996NewA....1..133D,1997ApJS..111...73K,1998NewA....3..687M,2000ApJS..128..561B,2000NewA....5..163B,2000ApJ...536L..39D,2001MNRAS.325..845K,2001NewA....6...79S,2001ApJ...550L.143K,2002NewA....7..373M,2002JCoPh.179...27D,2002JApA...23..185B,2003NewA....8..665B,2003PASJ...55.1163M,2003ApJS..145....1B,2004astro.ph..5220R,2004NewA....9..111D,2004PASJ...56..521M,2005MNRAS.364.1105S,2005NewA...10..393M,2005PASJ...57..849Y,2004NewA....9..137W,2006CoPhC.174..540T,2008arXiv0802.3215K} 
Along the way, code developers have also successfully met the challenge posed
by the emergence of distributed parallel programming.

In modeling gravitational clustering, cosmological N-Body codes should ensure
the following:
\begin{itemize}
\item
The universe does not have a boundary.  
Therefore cosmological simulations need to be run with periodic boundary
conditions\footnote{An exception are simulations of large spherical volumes
  that do not suffer significant deformation during the course of evolution.}.
The simulation volume should be large enough for the effects of missing modes
to be small, \citep{2005MNRAS.358.1076B,2006MNRAS.370..993B,2006MNRAS.370..691P,2007JApA...28..117P,2008arXiv0804.1197B}.
\item
The mass of each particle in simulations should be much smaller than mass
scales of interest in the simulation output. 
This is to ensure adequate mass resolution.
\item
Each particle in an N-Body simulation represents a very large number of
particles/objects in the universe.  
Thus we must ensure that pair-wise interaction of N-Body particles is softened
at scales comparable with the local inter-particle separation. 
If this is not ensured then the resulting two body collisions introduce errors
in the resulting distribution of particles,
\citep{1998ApJ...497...38S,2002MNRAS.333..378B}. 
\end{itemize}
In spite of the vast improvement in computing power, simulators have often
had to compromise on one or more of these points. 
Often, errors also creep in due to the approximate methods used for solving
for force in simulations.
A large fraction of current cosmological N-Body codes suffer from
collisionality or force biasing.  
Force is biased when softening lengths $\epsilon$ are much larger than the
local mean inter-particle separation, $\barrij$. 
Whereas a complementary effect, collisions, occur whenever $\epsilon \ll
\barrij$. 
The reader is referred to \citet{2001MNRAS.324..273D} for a detailed
discussion on these two effects.  
Codes that adapt their softening lengths to local densities generally are
mostly of the adaptive mesh refinement type.
These use a grid for solving the Poisson equation and often have anisotropies
in force at the scales comparable to the size of a grid cell.
Codes which use fixed softening lengths are not entirely collisionless, and,
in highly over-dense regions biasing of force also exists. 
The TPM,
\citep{1995ApJS...98..355X,2000ApJS..128..561B,2003ApJS..145....1B} 
is a particle based code with a one step adaptive resolution.  
However, in this case the use of the unmodified PM approach for computing the
long range force introduces significant errors at scales comparable to the
grid. 
In this work we describe a code which addresses all three issues of force
anisotropy, collisionality and force biasing by employing an adaptive softening
formalism with the hybrid TreePM code.  

The choice of the optimal softening length has been discussed at length 
\citep{1996AJ....111.2462M,2000MNRAS.314..475A,2001MNRAS.324..273D,2007MNRAS.374.1347P}. 
These studies were carried out in the context of isolated haloes in dynamical
equilibrium, e.g., Plummer and Hernquist profiles, therefore 
errors could be defined clearly.
It is also possible to compute and compare various physical quantities with 
analytical expressions derived from the distribution function. 
\citet{2001MNRAS.324..273D} derived analytical expressions for 
errors in the context of these profiles. 
This work suggested that the optimal softening length must adapt to the local
inter-particle separation as a function of space and time.
\citet{2007MNRAS.374.1347P} have developed an energy-momentum
conserving formalism with adaptive softening and demonstrated that it
was superior to fixed softening.  

The evolution of perturbations at small scales depends strongly on the mass
and force resolution. 
High force resolutions can lead to better modeling of dense haloes, but
gives rise to two body collisions in regions where the softening length is
smaller than the local inter particle separation
\citep{1998ApJ...497...38S,2002MNRAS.333..378B,2004MNRAS.348..977D,2004MNRAS.350..939B,2006MNRAS.370.1247E,2008ApJ...686....1R}. 
As all particles in very high density regions go through such a phase during
evolution, any errors arising due to two body collisions can potentially
effect the structure of high density haloes that form. 
A high force resolution without a corresponding mass resolution can also give
misleading results as we cannot probe {\it shapes}\/ of collapsed objects
\cite{1996ApJ...470L..41K}.
In addition, discreteness and stochasticity also limit our ability to measure
physical quantities in simulations, and these too need to be understood
properly \citep{2008MNRAS.387..397T,2008ApJ...686....1R}. 
In all such cases, the errors in modeling is large at small scales. 
It is important to understand how such errors may spread to larger scales and
affect physical quantities. 

This work is organized as follows. In \S{2} we describe the 
formalism for adaptive softening in a cosmological \N-body code. 
In \S{3} we review the TreePM method, 
which forms the backbone of our gravity solver and
describe how adaptive softening is implemented with it. 
We briefly discuss the performance characteristics of ATreePM in \S{4}. 
We discuss validation of the ATreePM code in \S{5} and we conclude in \S{6}. 

\section{Adaptive Force Softening}

\subsection{Formalism}

The aim of any collisionless \N-body code is to 
self consistently evolve the phase-space distribution function (DF)
$f(\br,\bv,t)$ under its own gravitational force field $\bF(\br,t)$:
\baq
\partial_t f &+& (\partial_{\br} f) . \bv + (\partial_{\bv} f) . \bF = 0\\
  \bF(\br,t) &=&   -G \int\int\dr^{\prime}\,\,\dv
  {\br-\br^\prime\over|\br-\br^\prime|^3}\,f(\br^\prime,\bv,t)
\eaq
The approach that one takes is to sample the DF, by \N \  phase-space points,
$\{\br_i,\bv_i\}_{_{t=t_i}} $, at initial time $t=t_i$. 
Liouville's theorem then states that evolving the trajectories of
these points to any time $t > t_i$ will be a representation of the DF
at that time.  
Since the system is a collisionless one, one has to suppress
artificial two-body collisions arising out of interactions between
particles which were generated for sampling the density field. 
One therefore assigns a finite size to N-Body particles which ensures
softening of force at small scales, instead of assuming these to be
point particles. The density field when sampled by point particles,
\beq 
\rho({\mathbf{r}}) = \int\dv f(\br,\bv,t) \equiv
\sum_{j=1}^N m_j \delta_D^3({\mathbf{r}} -{\mathbf{r_j}})
\eeq
is now smoothed at small scales if we assign a finite size $\epsilon$
to every particle:
\beq
\rho({\mathbf{r}}) = \int\dv f(\br,\bv,t) \equiv
\sum_{j=1}^N m_j W(|{\mathbf{r}} -{\mathbf{r_j}}|,\epsilon)
\eeq
Where $W(u,\epsilon)$ is known as the smoothing kernel and we have assumed
that particles are spherical in shape.
Here $m_j$ is the mass of the $j^{th}$ particle.
We can now integrate the Poisson equation to obtain the expression for
the kernel for computing force and potential.
Both the quantities are softened at scales below the softening length
$\epsilon$.  
We choose to work with the cubic spline kernel \citep{1985A&A...149..135M}
whose expression is given below.
Complete expressions for the potential and force are given in the Appendix
(See Eqn.(\ref{pot_fixedh}, \ref{force_fixedh})).
\beq
 W(u,\epsilon) = \frac{8}{\pi \epsilon^3}
  \left\{
    \begin{array}{ll}
      1 - 6(\frac{u}{\epsilon})^{^2} + 6(\frac{u}{\epsilon})^{^3} ,
      & \,\mbox{$\,\,\,0\,\,\leq \frac{u}{\epsilon} < 0.5$} \\
      2(1- \frac{u}{\epsilon})^{^3},
      & \,\mbox{$0.5 \leq \frac{u}{\epsilon} < 1.0$} \\
      0,
      & \,\mbox{$1.0 \leq \frac{u}{\epsilon}$}
    \end{array}
  \right.
\label{eq_mon_kernel}
\eeq
In the context of individual softening lengths the density at the location of
the $i^{th}$ particle is given by: 
\baq
\rho(r_i) = \sum_{j=1}^N m_j W(r_{ij},\epsilon_i) = 
\sum_{j=1}^{n_n} m_j W(r_{ij},\epsilon_i)
\eaq
where {\emph{i,j}}  indicate the particle indices,
$r_{ij}=|\br_i-\br_j|$ and $\epsilon_i\equiv\epsilon(\br_i)$. 
The summation in principle can be extended upto infinity if the kernel is
infinite in extent (e.g. plummer, gaussian kernels). 
But since such kernels tend to bias the force
\citep{2001MNRAS.324..273D,2007MNRAS.374.1347P} we will work only with those
kernels which have compact support, in particular the cubic spline kernel. 
Such kernels ensure that the force is Newtonian beyond the
softening scale. 
$n_n$ is the number of the nearest nearbors within $\epsilon_i$ for the
particle $i$. 
In the discussion that follows we assume that this number is fixed for
every particle and sets the value of the softening length $\epsilon_i$.
We implicitly assume it in our summation. 
Integrating the Poisson equation we obtain the Green's function for the
potential $\phi_{ij} \equiv  \phi(r_{ij},\epsilon_i)$, where the functional
form for $\phi$ is given in the Appendix (See Eqn.(\ref{pot_fixedh},
\ref{force_fixedh})).   

With the introduction of individual softening lengths for particles, the
symmetry of the potential is lost and  momentum conservation is violated. 
Since $\epsilon(\br) $ is now a local quantity, the EOM is incomplete if one
takes the expression of force with the fixed softening length $\epsilon$
replacing it by a local softening length $\epsilon(\br)$.  
Hence energy conservation also gets violated.
This is because the force is derived from the potential and with the
introduction of a local softening length $\epsilon(\br)$ in the potential, the
gradient must also act on $\epsilon(\br)$ giving us an extra term.  
Traditionally this \emph{grad-}$\epsilon$ ($\gradeps$) term has been
ignored since in typical applications these were found to be
subdominant when compared to the usual force 
\citep{1982JCoPh..46..429G,1988MNRAS.235..911E,1990ApJ...349..562H,1992ARA&A..30..543M}.
It has been shown recently \citep{2007MNRAS.374.1347P}, that this term
plays an important role in N-Body simulations.  
We study the impact of ignoring this term in Cosmological simulations.

A remedy for momentum non-conservation is to use a symmetrized
softening length  
\begin{equation}
  \epsilon_{ij} = \frac{1}{2}(\epsilon_i+\epsilon_j) 
\end{equation}
and plug it into the expression for density to re-derive a symmetrized
expression for the potential as $\phi(r_{ij},\epsilon_{ij})$. 
This prescription changes the softening length and hence the neighborlist,
which one has to recompute.  
Another disadvantage is that with this prescription for symmetrization, the
number of neighbors is not fixed for every particle and hence errors in all
smoothed estimates are not the same for every particle. 
An alternate method \citep{1989ApJS...70..419H} is to symmetrize the kernel
itself. 
\baq
\widetilde{W}_{ij} &=&  
\widetilde{W}_{ji} = \frac{1}{2}
\left[
W(r_{ij},\epsilon_j)+W(r_{ij},\epsilon_i)
\right] \nonumber \\
\widetilde{\phi}_{ij} &=& 
\frac{1}{2}
\left[
\phi(r_{ij},\epsilon_j) + \phi(r_{ij},\epsilon_i)
\right]
\eaq
The total potential is thus 
\beq
\Phi_{tot} = \frac{1}{2}\sum_{i,j}^N \widetilde{\phi}_{ij}
\eeq
We can use this to write a Lagrangian which is manifestly symmetric 
and this ensures momentum conservation.
Energy is conserved only if the $\nabla\epsilon$ term is retained in the EOM. 
\beq
L = \sum_i^N \frac{1}{2}m_iv_i^2 - \frac{G}{2}\sum_{i,j}^N m_im_j \widetilde{\phi}_{ij}
\eeq

The EOM of motion can be derived with this Lagrangian (the reader is 
referred to \citet{2007MNRAS.374.1347P} for details)
\baq
\frac{d{\bf v}_{i}}{dt}  = &-&G\sum_{j} m_{j}
\widetilde{\phi}^\prime_{ij}
\frac{\br_{i} - \br_{j}}{\vert \br_{i} - \br_{j} \vert}  \nonumber \\
 &-&\frac{G}{2} \sum_{j} m_{j} \left[
  \frac{\zeta_i}{\Omega_i}
\pder{W_{ij} (\epsilon_{i})}{\br_{i}} +  
  \frac{\zeta_j}{\Omega_j}
\pder{W_{ij} (\epsilon_{j})}{\br_{i}}
\right]
\eaq
Where the first term is the standard Newtonian force term (we refer to it as
the $\nabla\phi$ term). 
The second is the energy conserving $\nabla\epsilon$ term which would be zero
for fixed softening. 
Notice that all terms are antisymmetric in $i,j$ and hence the total momentum
is conserved. 
Here $\zeta$ and $\Omega$ are:
\baq
\zeta_{i} \equiv \pder{\epsilon_{i}}{\rho_{i}}
\sum_{j} m_{j} \pder{\phi(r_{ij},\epsilon_i)}{\epsilon_i}\\
\Omega_i = \left[1 - \pder{\epsilon_i}{\rho_i}\sum_{j} m_{j} 
\pder{W_{ij}(\epsilon_i)}{\epsilon_i}\right]
\eaq
Expressions for $\pder{W}{\br}$, $\pder{W}{\epsilon}$ and 
$\pder{\phi}{\epsilon}$ are given in the 
Appendix (See \ref{kernel_dwde},\ref{kernel_dwdr} and\ref{pot_dphide}).
As $\epsilon_i \propto \rho_i^{-\frac{1}{3}}$ 
we have $\pder{\epsilon_{i}}{\rho_{i}} = -\frac{\epsilon_i}{3\rho_i}$.
The term $\Omega$ term ensures that the EOM is accurate to all orders
in $\epsilon$ \citep{2002MNRAS.333..649S}.

\section{The Adaptive TreePM Method}

\subsection{The TreePM Method}

The TreePM algorithm \citep{2002JApA...23..185B,2003NewA....8..665B} is a
hybrid N-Body method which improves the accuracy and performance of
the Barnes-Hut (BH) Tree method \citep{1986Natur.324..446B} by
combining it with the PM method 
\citep{1997Prama..49..161B,2005NewA...10..393M,1983MNRAS.204..891K,1983ApJ...270..390M,1985ApJ...299....1B,1985A&A...144..413B,1988csup.book.....H}. 
The TreePM method explicitly breaks the potential into a short-range and a
longe-range component at a scale $r_s$. 
The PM method is used to calculate long-range force and the short-range force
is computed using the BH Tree method.  
Use of the BH Tree for short-range force calculation enhances the force 
resolution as compared to the PM method. 

The gravitational force is divided into a long range and a short range part
using partitioning of unity in the Poisson equation.
\begin{eqnarray}
\varphi_k &=& -\frac{4\pi G\rho_k}{k^2} \nonumber\\
\varphi_k &=&  \varphi_k^{lr} + \varphi_k^{sr} \nonumber \\
\varphi_k^{lr} &=& -\frac{4\pi G\rho_k}{k^2} \exp \left(-k^2r_s^2\right) \\
\varphi_k^{sr} &=& -\frac{4\pi G\rho_k}{k^2} 
\left[1- \exp \left(-k^2r_s^2\right)\right] 
\end{eqnarray}
Here $\varphi_k^{sr}$ and $\varphi_k^{lr}$ are the short-range and long-range
potentials in Fourier space. $\rho$ is the density, G is the gravitational 
coupling constant and $r_s$ is the scale at which the splitting of the 
potential is done. 
It has been shown that this particular split between the long range
and the short range force is optimal amongst a large class of suitable
functional forms \citep{2003NewA....8..665B}.
The long-range force is computed in Fourier space with the PM method and the
short-range force is computed in real space with the Tree method.  
The short range potential and force in real space are:
\baq
\varphi^{sr}(\br,\epsilon) &=& 
Gm\phi(\br,\epsilon)\mbox{erfc}\left(\frac{r}{2r_s}\right) \\
\bF^{sr}(\br,\epsilon) 
&=& Gm\mathbf{f}(\br,\epsilon)C\left(\frac{r}{r_s}\right)\\
C\left(\frac{r}{r_s}\right) &=& 
\left[
  \mbox{erfc}\left(\frac{r}{2r_s}\right) + \frac{r}{r_s\sqrt{\pi}}
  \exp\left(-\frac{r^2}{4r^2_s} \right)
\right]  
\eaq
Here $\mbox{erfc}$ is the complementary error function. 
$\phi(\br,\epsilon)$ and $\mathbf{f}(\br,\epsilon)$ are the usual potential
and force kernels, respectively. 
$C(r/r_s)$ modifies the softened Newtonian force kernel
$\mathbf{f}(\br,\epsilon)$ to the short-range force $\bF^{sr}(\br,\epsilon)$.  
The expression for $\phi(\br,\epsilon)$ and $\mathbf{f}(\br,\epsilon)$
depend on the kernel $W(r,\epsilon)$ used for smoothing and are given for
cubic spline in the Appendix.  
We find that tabulating $\mbox{erfc}(u)$ and $C(u)$ 
and using interpolation to compute these functions is much more 
effective than calculating them every time.
The short range force is below $1\%$ of the total force at $r \geq 5 r_s$. 
The short range force is therefore computed within a sphere of radius $r_{cut}
\simeq  5 r_s$ using the BH tree method.  

The long range force falls below $1\%$ of the total force for $r \leq
0.5 r_s$.   
Thus the choice of softening length does not affect the long range
force in a significant manner as long as the force softening is done
with a kernel that has a compact support and the softening length is
below $0.5 r_s$. 

The BH tree structure is built out of cells and particles.  
Cells may contain smaller cells (sub-cells) within them.  
Sub-cells can have even smaller cells within them, or they can contain a
particle. 
In three dimensions, each cubic cell is divided into eight cubic
sub-cells.  
Cells, as structures, have attributes like total mass, location of center of
mass and pointers to sub-cells.  
Particles, on the other hand have the usual attributes: position, velocity and
mass. 

Force on a particle is computed by adding contribution of other
particles or of cells.  
A cell that is sufficiently far away can be considered as a single entity and
we can add the force due to the total mass contained in the cell from its
center of mass.  
If the cell is not sufficiently far away then we must consider its
constituents, sub-cells and particles.  
Whether a cell can be accepted as a single entity for force calculation is
decided by the cell acceptance criterion (CAC).  
We compute the ratio of the size of the cell $L_{cell}$ and the distance $r$
from the particle in question to its center of mass and compare it with a
threshold value 
\begin{equation}
\theta = \frac{L_{cell}}{r} \leq \theta_c  
\end{equation}
The error in force increases with $\theta_c$.  
Poor choice of $\theta_c$ can lead to significant errors
\citep{1994JCoPh.111..136S}.  
Many different approaches have been tried for the CAC in order to minimize
error as well as CPU time usage
\citep{1994JCoPh.111..136S,2001NewA....6...79S}.  
The tree code gains over direct summation as the number of contributions to
the force is much smaller than the number of particles.
 
The TreePM method is characterized therefore by three parameters, 
$r_s$,$r_{cut}$ and $\theta_c$. 
For a discussion on the optimum choice of these parameters the reader is
referred to \citet{2003NewA....8..665B}. 
For all our tests we choose conservative values $r_s = 1.0$, $r_{cut} =
5.2r_s$ and $\theta_c = 0.3$ which give errors below $1\%$ in force. 
All lengths are specified in units of the PM grid. 

\subsection{The Adaptive TreePM Method}

We choose to implement adaptive softening with a  modified TreePM code 
\citep{2008arXiv0802.3215K}, which incorporates Barnes optimization 
using \emph{groups} 
\citep{1990JCoPh..87..161B,1991PASJ...43..621M,2005PASJ...57..849Y}, 
into the TreePM code. 
In principle one can also incorporate a similar formalism for
treecodes \citep{2001NewA....6...79S}, \p3m codes
\citep{1991ApJ...368L..23C,1995ApJ...452..797C} and other variants like  
TPM \citep{1995ApJS...98..355X}, TreePM
\citep{2002JApA...23..185B,2005MNRAS.364.1105S} and
GOTPM \citep{2004NewA....9..111D}. 
We shall discuss one advantage of using the TreePM code with the group
optimizations below.

\subsubsection{Estimating the Softening Length}

Our first task is to get an estimate of the softening length for 
each particle. 
A natural way to extract a local length scale uses the numerical value
of local density. 
The local number density is related to the softening length as:
\beq
 \epsofr \sim \left(\frac{n_{_n}}{n(\br))}\right)^{\frac{1}{3}} \label{eps_con}
\eeq
$n(\br)$ is the number density at the location of the particle and we
assume all particles have the same mass. 
Here, $n_n$ is a reference number and we take it to be the number of
neighbors used for estimation of the number density.
The above equation is implicit and can be solved iteratively, see, e.g.,  
\citet{2002MNRAS.333..649S,2005MNRAS.364.1105S,2007MNRAS.374.1347P} for
details. 
\citet{2007MNRAS.374.1347P} have shown that errors are not very
sensitive to the exact value of $n_n$. 
We choose $n_n = 32$ in our simulations, and also comment on variation
in results with this choice.

We are using the formalism developed by \citet{2007MNRAS.374.1347P} for
achieving an adaptive resolution in gravitational interactions of
particles. 
As the formalism was developed in the context of SPH codes, and some
of the quantities required can be computed naturally using the SPH
method, the same was used even though the gravitational interaction is
completely collisionless. 
We follow a similar implementation and use methods commonly used in
SPH simulations, even though there are no hydrodynamical effects present in
the gravitational interaction being studied here.
For an overview of SPH methods, please see 
\citet{1977AJ.....82.1013L,1977MNRAS.181..375G}.
The SPH methods assign values for functions like the density to
particles by averaging over nearest neighbors.
The list of nearest neighbors is an essential requirement for
computing anything using these methods. 
All quantities ($\rho$, $\nabla \epsilon$, $\Omega$, $\zeta$ etc.) can
be computed at runtime with this neighborlist once we have converged
to a value for $\epsilon$ by solving Eqn.(\ref{eps_con}).
We compute the neighborlist using linked lists \citep{1989ApJS...70..419H}. 

We put bounds on the maximum and minimum softening 
lengths, $\epsmax$ and $\epsmin$. 
A maximum bound is required so that force softening 
is restricted to the short-range force only, we choose $\epsmax =
\frac{r_s}{2}$ in order to ensure that the long range force is of
order $1\%$ of the total force (or smaller) at scales where force
softening is important. 
This ensures that any errors arising from non-modification of the long
range force are smaller than $1\%$, if one puts a lower threshold on
the maximum allowed error then the scale $\epsmax$ has to be lowered
correspondingly. 
Alternatively, one can work with a larger $r_s$ and then it becomes possible
to allow a larger $\epsmax$.
A lower bound is also required for the reason that we do not
want a few isolated highly over-dense regions to dominate the CPU time
requirements.  
The value of the lower bound must correspond to densities that are
much higher than highest over-densities of interest in the simulation.
We set this lower bound $\epsmin = {r_s}/{1000}$, which
corresponds to $\rho\sim 10^{10} \bar\rho$.
The lower bound however is not critical to the structure of the code and may
be omitted.

In our implementation a neighbor search is carried out only
upto $\epsmax$. 
Particles which do not have $n_n$ neighbors within $\epsmax$ are
assigned $\epsilon_i=\epsmax$ and the spherical
top hat (STH) density is assigned with the number of neighbors within
$\epsmax$. 
For these particles we assign $\Omega = 1$ and $\zeta=0$, which makes
their $\gradeps_i = 0$. 
A similar assignment is carried out for particles which have
$\epsilon_i \leq \epsmin$.
The $\gradeps$ term is calculated before the short-range force so that
the individual softening lengths needed for short-range force
calculation are also assigned in the process. 

\subsubsection{Memory Requirements}

Even though we use two separate data structures, namely 
linked lists for $\gradeps$ and tree for $\gradphi$ in order to
compute the total short-range force, additional memory requirements
compared to TreePM are minimal: we require one additional array for
storing the softening lengths. 
The $\gradeps$ term does not require an additional array since it is a
component of the short-range force and it can be computed at run
time. 
This is because the two data structures are never required at the same
time.  
We require specification of the largest force softening length in a given
cell (See the subsection on the cell acceptance criterion below.).
This amounts to a single precision array of the same size as the number of
cells in the tree.

An advantage of using an analytical splitting of force, 
in the manner TreePM does, is that computation 
of short-range force does not need global data structures.
For example one can geometrically divide regions into smaller regions 
and construct local trees and linked lists in them 
(just like one would go about doing it in a distributed code) and 
iterate through these regions for computing short-range force 
instead of constructing one global
data structure for the entire simulation volume for computing the
short-range force \citep{2004NewA....9..111D}. 
This reduces memory usage significantly and the dominant part is taken up
by the arrays required for computing the long range PM force.

\subsubsection{Timestepping Criterion and Integration}

We have implemented a hierarchical time integrator similar to that used in
GADGET-2 \cite{2005MNRAS.364.1105S}, in which particle trajectories are
integrated with individual timesteps and synchronized with the largest
timestep. 
As we allow the block time step\footnote{Same as the largest time step.} to
vary with time, we work with the so called KDK approach (Kick-Drift-Kick) in
which velocities are updated in two half steps whereas position is updated in a
full step.
It can be shown that with a variable time step, KDK performs better than DKD
(Drift-Kick-Drift) (see \citet{2005MNRAS.364.1105S} for
details.). 
We give separate PM (long range, global) kicks  and Tree (short range,
individual) kicks\footnote{Tree kick includes the contribution of the
  $\gradeps$ term.}.   
The block timestep $\Delta t^{^{PM}}$ is determined by the particle
which has the maximum PM acceleration $a^{^{PM}}_{max}$:
\baq
\Delta t^{^{PM}} = \delta t 
\left(
\frac{\epsilon_{max}}{a^{^{PM}}_{max}}
\right)^{1/2}
\eaq
Here $\delta t$ is the dimensionless accuracy parameter. 
In our implementation of the hierarchy of time steps, the smaller time steps
differ by an integer power ($n$) of $2$ from the largest, block time step. 
An array is then used to store the value $n$ which determines the timestep of
the particle. Individual timesteps $\Delta t_i$ 
are first calculated:
\baq
\Delta t_i^{sr} = \delta t 
\left(
\frac{\epsilon_i}{a_i^{sr}}
\right)^{1/2}
\label{eq_tstep_atpm_sr}
\eaq
and then the appropriate hierarchy $n$ is chosen depending 
on this value. Here $a^{sr}_i$ and $\epsilon_i$ are the modulus of the 
individual short-range acceleration (sum of the $\gradphi$ and
$\gradeps$ terms) and the softening lengths respectively.
TreePM has a similar time-stepping criterion with $\epsilon_i$ replace 
by $\epsilon$.

The code drifts all the particles with the smallest timestep to the next time,
where a force computation is done for particles that require an updation of
velocity (Kick). 
However the neighborlist and individual softening lengths $\epsilon_i$ are
computed for all particles at every small timestep. 
This is because, even though some particles do not require a velocity update,
their neighbors might require one for which they would contribute through
their updated softening lengths.  
The $\gradeps$ term however can be computed only for those particles requiring
a velocity update. 

Within a given block time step, the smaller time steps are constant for a
given particle.  
The time step changes across block time step and this brings in inaccuracies
in evolution of trajectories. 
It is possible, in principle, to ensure that the second order accuracy
is maintained here.
However, we find that the time steps for particles change very slowly and this
change does not affect trajectories in a significant manner. 

The Courant condition is satisfied for the choice of $\delta t$ we use.
Indeed, we chose $\delta t$ by requiring that two particles in a highly
eccentric orbit around each other maintain the trajectory correctly for tens
of orbits.

\begin{figure*}
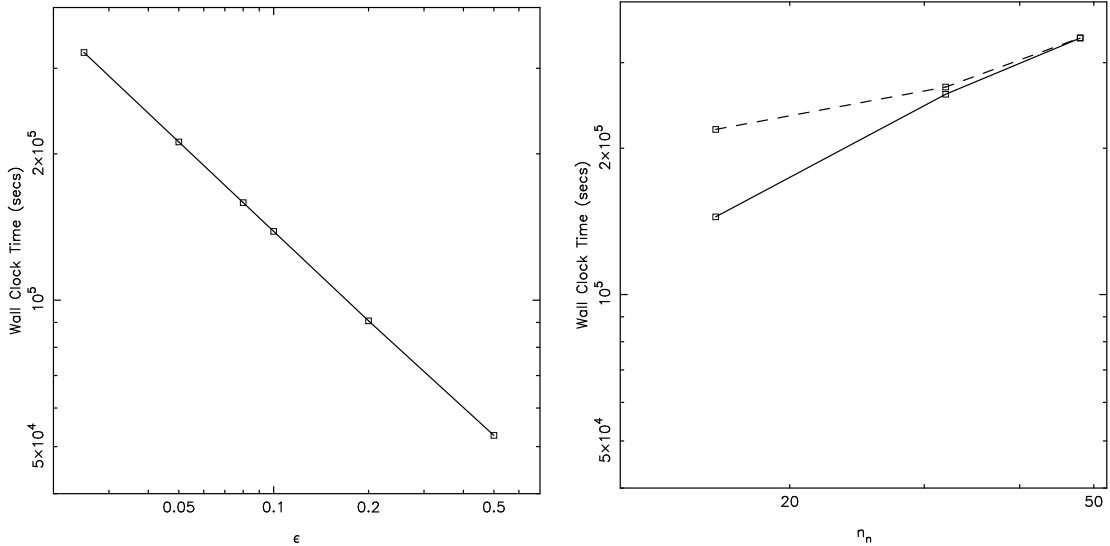

\begin{center}
\begin{tabular}{cc}
\includegraphics[width=2.8truein]{1a.ps} &
\includegraphics[width=2.8truein]{1b.ps} \\
\end{tabular}
\end{center}
\caption{Wall clock timing for TreePM (left) as a function of $\epsilon$ and 
ATreePM (right) with $\gradeps$ (solid) and 
without $\gradeps$ (dashed) as a function of $n_n$}
\label{fig_timing}
\end{figure*}

\subsubsection{Cell Acceptance Criterion For ATreePM}

The adaptive force resolution formalism requires us to symmetrize
force between particles that are separated by a distance smaller than
the larger of the two softening lengths. 
Without this, the momentum conservation cannot be ensured.
For pairs of particles separated by larger distances, there is no need
to explicitly symmetrize force as there is no dependence on the
softening length at these scales\footnote{As an aside we would like to
  note that the Tree method does not conserve momentum explicitly.
This is because the tree traversal approximates the force due to
pairwise interactions, and in the process the pairwise symmetry is
lost. 
In the modified Tree method
\citep{1990JCoPh..87..161B,1991PASJ...43..621M,2005PASJ...57..849Y,2008arXiv0802.3215K} 
explicit pairwise force is computed for particles within each \emph{group}. 
Particles within a group have a common interaction list for force due
to particles outside the group. 
Exact pairwise PP force is computed explicitly for intra-group
particles and we have a pairwise symmetry for this component, but for
interaction with particles outside the group there is no explicit
pairwise symmetry and hence no explicit momentum conservation.}.
Thus the cell acceptance criterion needs to be changed within the tree
part of the code to ensure that for pairs of particles separated by
the critical distance, forces are computed through pairwise
particle-particle (PP) interaction.
In our implementation, along with particles, cells (and hence groups)
are assigned softening lengths corresponding to the particle contained
within the cell that has the largest softening length. 
The CAC, that is modified when we go from the TreePM to the modified
TreePM \citep{2008arXiv0802.3215K}
has to be changed again to take
into account the largest softening lengths for particles within cells
and groups whose interaction is to be computed.
\beq
\frac
{(L_{_{cell}} + \epsilon_{_{cell}}^{^{max}})} 
{(r_{g_{cm}}-L_{group}-\epsilon_{_{group}}^{^{max}})} \leq \theta_c
\eeq
$\epsilon_{_{cell}}^{^{max}}, \epsilon_{_{group}}^{^{max}}$ are the 
softening lengths of the cell and group (in the sense explained above)
respectively. 
$r_{g_{cm}}$ is the distance separating the centers of mass of the
group and the cell. 
$L_{_{cell}}$ is the size of the cell and $L_{_{group}}$ is the
distance to the furthest particle from the center of mass of the
group. 
This CAC ensures that the interaction of particles separated by less
than the softening length is computed in a direct pairwise manner and
hence can be explicitly symmetrized.
These are also the pairs for which the $\gradeps$ term needs to be
computed. 

\section{Performance characteristics}

\subsection{Timing}

We now look at the wall clock time as a measure of performance 
between different codes. 
We studied evolution of a power law model with $n=-1$ using $128^3$ particles
up to the stage where the scale of non-linearity is $6$ grid lengths. 
More details of the run are given in the section on validation of the ATreePM
code. 
Figure~\ref{fig_timing} shows wall clock time as a function of softening
length $\epsilon$ for TreePM (left panel) and the wall clock time as a
function of $n_n$ for the ATreePM codes. 

We can qualitatively understand the slope for the TreePM curve by looking at
the equivalent time-stepping criterion as Eqn.(\ref{eq_tstep_atpm_sr}) for
TreePM.
\beq
\Delta t_i^{sr} = \delta t 
\left(
\frac{\epsilon}{a_i^{sr}}
\right)^{1/2}
\label{eq_tstep_tpm_sr}
\eeq
From here the naive expectation is that the time taken should scale as
$\epsilon^{-0.5}$. 
The slope of the curve is in the range $-0.55$ to $-0.65$. 
The reason for this small deviation lies in our use of a hierarchy of time
steps, where trajectories of all the particles are not updated at every time
step. 
However, positions of particles are updated at every time step and this
operation as well as those related to creating the tree structure at every
step add an overhead. 
This overhead becomes more and more important at small $\epsilon$ where we
have many more levels of hierarchy realized in a simulation. 
This leads to steepening of the curve from the simple expectation given
above.

\begin{figure*}
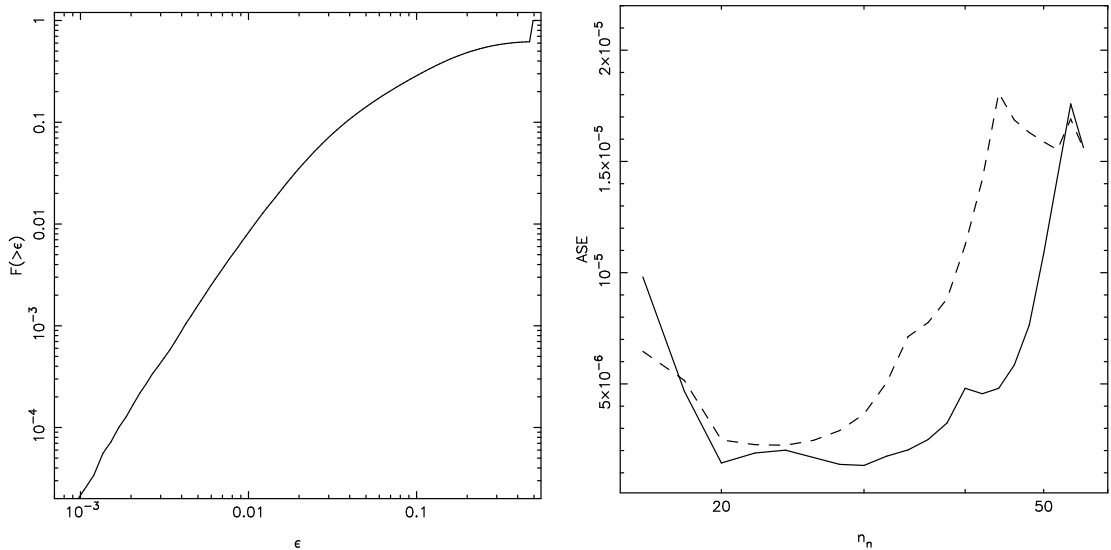

\begin{center}
\begin{tabular}{cc}
\includegraphics[width=2.8truein]{2a.ps} &
\includegraphics[width=2.8truein]{2b.ps} 
\end{tabular}
\end{center}
\caption{\emph{Left}: Cumulative distribution of softening lengths at the
  final epoch of an \N-Body simulation (See Figure~\ref{fig_slice_full} bottom-left). 
  \emph{Right}: Average square error ($ASE$) for ATreePM with 
  $\gradeps$ (solid line) and ATreePM without $\gradeps$ (dashed line) 
  as a function of $n_n$}.
\label{fig_err}
\end{figure*}

For the ATreePM, we expect softening lengths to be larger for larger $n_n$. 
On the other hand, a larger $n_n$ implies a larger neighborlist and the time
taken for setting up the neighborlist increases. 
The second effect is the dominant one and we see that the time taken for
Adaptive TreePM increases with $n_n$. 
We see that time taken by both variants of ATreePM is similar for $n_n=32$ and
$48$.  
This indicates that the time taken for calculation of the $\gradeps$ term is
negligible. 
There is a difference between the timing for $n_n=16$, as the code with the
$\gradeps$ does not evolve the system correctly: this can be seen in all the
indicators like the amplitude of clustering, mass function, etc., presented in
the next section. 
We see that TreePM with $\epsilon=0.025$ takes $50\%$ more time than ATreePM
with $n_n=32$, whereas the time taken are comparable with ATreePM with
$n_n=48$.  

Since $\epsilon$ is assigned by hand in TreePM, the use of a small 
softening length means a considerable number of particles 
obtain a small timestep even though the acceleration for these particles is
small and does not vary rapidly with time.
In a hierarchical integrator that we use, this means that force computation 
is done more often within a block timestep. 
In ATreePM on the other hand a small softening is only assigned to particles 
in over-dense regions.
This saves time in the under-dense and not so over-dense regions and the
ATreePM code devotes more time to evolving trajectories in highly over-dense
regions. 
This is illustrated in Figure~\ref{fig_err} 
where we have plotted the cumulative
distribution of softening lengths at the final epoch in one of the simulations
used here (see lower-left panel of Figure~\ref{fig_slice_full}).
The softening length was computed here for $n_n=32$. 
We find that only $5\%$ of the particles have a softening length smaller than
the smallest softening length $\epsilon=0.025$ used for the fixed resolution
simulations.  
Even at this epoch where highly non-linear clustering is seen, nearly half the
particles have a softening length corresponding to the maximum value of $0.5$.
This shows how we are able to evolve the system in an ATreePM simulation with
a lower computational cost while resolving highly over-dense regions.

\subsection{Errors}

In this section we discuss the dependence of errors on $n_n$.
In cosmological simulations it is difficult to define errors when softening
lengths are varied due to the lack of a reference setup.  
Nevertheless we can choose the optimal softening length such that one
minimizes the globally averaged fluctuation in force as the
softening lengths are varied.   
Following \citet{2007MNRAS.374.1347P}, we define average square error: 
\beq
ASE(n_n) = \frac{C}{N}\sum_i^N
\left|
\mathbf{f}_i(n_n)-\mathbf{f}_i(n_n+\Delta n_n)
\right|^2
\eeq
$N$ is the total number of particles. 
$C$ is a normalization constant which is taken as \(1/f_{max}^2\), 
with $f_{max}$ being the largest value of force in either runs.
$\Delta n_n$ is the change in $n_n$ and we choose this to be $8$.
With other values of $\Delta n_n$ the overall behaviour of the error
plot remains qualitatively the same. 
Figure~\ref{fig_err} shows $ASE$ as function of $n_n$ for ATreePM
with $\gradeps$ (solid line) and ATreePM without $\gradeps$ (dashed line).  
These were computed for the same clustered distribution of particles. 
The qualitative behavior of $ASE$ here is same as seen in
\citep{2007MNRAS.374.1347P}.  
At small $n_n$ ATreePM with $\gradeps$ has larger errors than ATreePM without
$\gradeps$.  
This is also reflected in the poor evolution of density fluctuations with
$n_n=16$ for the ATreePM with $\gradeps$, discussed in the next section. 
Both variants have a minima, the value of error at the minima being lower for
ATreePM with the $\gradeps$ term. 
Another interesting feature is that the region where errors are small is
fairly broad for the ATreePM with the $\gradeps$ term.
For larger $n_n$ the error increases sharply for ATreePM without the
$\gradeps$ term. 
Thus the optimal configuration is the one with the $\gradeps$ term and $20
\leq n_n \leq 32$.

\begin{figure*}
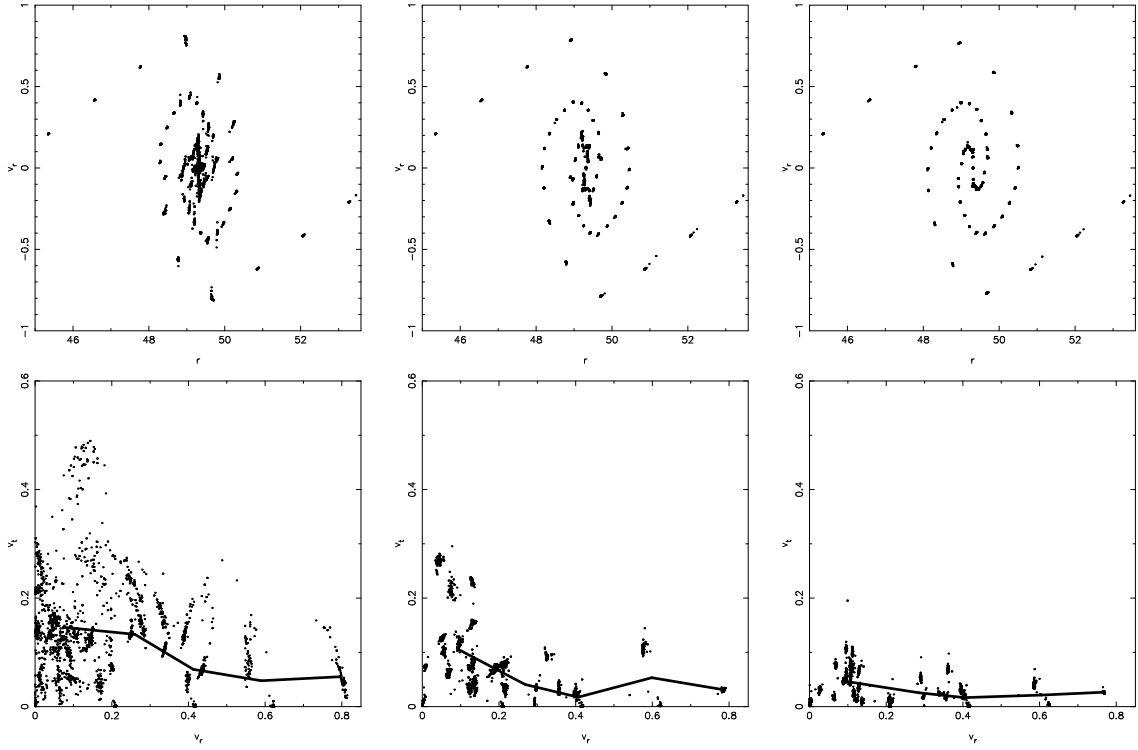

\begin{center}
\begin{tabular}{ccc}
\includegraphics[width=1.86truein]{3a.ps} &
\includegraphics[width=1.86truein]{3b.ps} &
\includegraphics[width=1.86truein]{3c.ps} \\
\includegraphics[width=1.86truein]{3d.ps} &
\includegraphics[width=1.86truein]{3e.ps} &
\includegraphics[width=1.86truein]{3f.ps} \\
\end{tabular}
\end{center}
\caption{Top row: Phase plots of the normal (to one of the planes of
  collapse) component of the velocity vs normal component of particle 
  displacement for TreePM with $\epsilon = \frac{r_s}{10}, 
  \frac{r_s}{4}, \frac{r_s}{2} $ (columns 1-3 respectively). 
  Bottom row: Transverse component of velocity vs Normal component of velocity 
  along one of the planes of collapse for the above runs. The scatter
  plot shows this for a random subset of particles, whereas the line
  shows the average value in a few bins.}
\label{fig_plwave_treepm}
\end{figure*}
\begin{figure*}
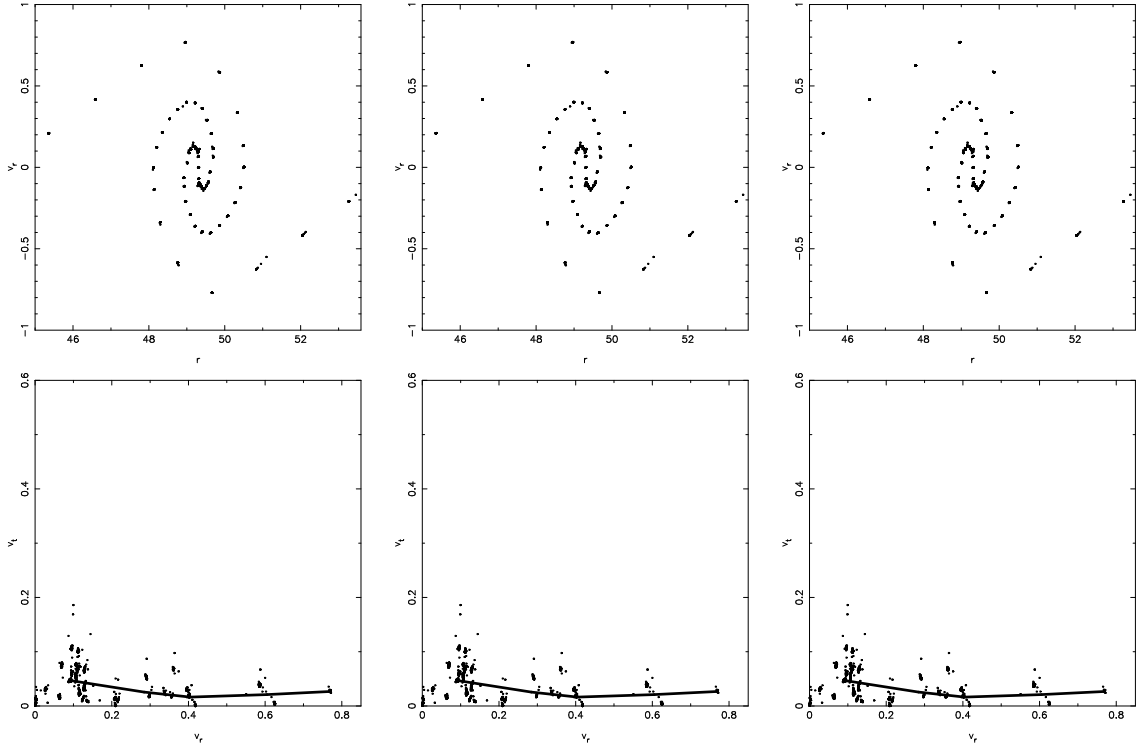

\begin{center}
\begin{tabular}{ccc}
\includegraphics[width=1.86truein]{4a.ps} &
\includegraphics[width=1.86truein]{4b.ps} &
\includegraphics[width=1.86truein]{4c.ps} \\
\includegraphics[width=1.86truein]{4d.ps} &
\includegraphics[width=1.86truein]{4e.ps} &
\includegraphics[width=1.86truein]{4f.ps} \\
\end{tabular}
\end{center}
\caption{Top row: Phase plots of the normal 
  component of particle displacement vs normal component of 
  the velocity for ATreePM with 
  $ \epsilon_i \leq \frac{r_s}{2}$ 
  for $n_n=16$ without $\gradeps$,    $n_n=16$ with  $\gradeps$ and
  $n_n=32$ with  $\gradeps$ 
  (columns 1-3 respectively).
  Bottom row: Normal component of velocity vs Transverse component of velocity 
  along one of the planes of collapse for the above runs. The scatter
  plot shows this for a random subset of particles, whereas the line
  shows the average value in a few bins.}
\label{fig_plwave_atreepm}
\end{figure*}

\section{Validation of the Adaptive TreePM code}

Cosmological N-Body simulations lack the equivalent of equilibrium
distribution functions for haloes, e.g., Plummer halo that may be used
to validate a new code.  
Use of such equilibrium distributions allows one to quantify errors in
a clean manner. 
However, given that the formalism we use is the same as that presented
by \citet{2007MNRAS.374.1347P}, and that their implementation works
well with such tests gives us confidence in the formalism.
In the following discussion, we will test the Adaptive TreePM code in
a variety of ways and look for numerical convergence. 
We compare the performance of the Adaptive TreePM with the fixed
resolution TreePM.
We also study the role of the $\gradeps$ term and check if dropping 
this term leads to any degradation in performance of the code. 
This last point is important as most AMR codes do not have an
equivalent term in the equation of motion.


\subsection{Two-body collisions: plane wave collapse}

We start by checking whether the adaptive force softening suppresses
two body collisions in the Adaptive TreePM code. 
We repeat a test recommended by \citet{1997ApJ...479L..79M} where
they study the collapse of an oblique plane wave. 
In this test, the collapse should not lead to any transverse motions
if the evolution is collisionless. 
\citet{1997ApJ...479L..79M} had shown that codes where the force
softening length is much smaller than the inter-particle separation
are collisional and lead to generation of significant transverse
motions. 
Some authors \citep{2005ApJS..160...28H} have stated that the failure to
ensure planar symmetry may not have a one to one correspondence with
collisionality. 
However, the test is nevertheless an important one and we present the results
with the ATreePM code here. 

We use exactly the same initial perturbations as used by
\citet{1997ApJ...479L..79M}. 
The simulations are done with $64^3$ particles and a $64^3$ grid.
We choose $r_s=1$ grid length.  
Other simulation parameters are as described in the subsection \S{3.1}
on the TreePM code.
The output is studied at $a=3$, following \citet{1997ApJ...479L..79M}. 

We first conduct the test with the fixed resolution TreePM code. 
Top row of the Figure~\ref{fig_plwave_treepm} shows the phase portrait
along the direction of collapse for different choices of the force
softening length. 
The softening length varies between $0.1 \leq \epsilon \leq 0.5$ in
units of the mean inter-particle separation. 
We see that the phase portrait in the multi-stream region is heavily
distorted for the smallest force softening length but is correct for
the largest force softening length used here. 
This reinforces the conclusions of \citet{1997ApJ...479L..79M} that
using a force softening length that is much smaller that the mean
inter-particle separation leads to two body collisions.
This point is presented again in the lower row of
Figure~\ref{fig_plwave_treepm} where we show a scatter plot of
modulus of the transverse (to the direction of collapse) velocities
and radial velocities. 
This is shown for a random sub-set of all particles.
We see that transverse motions are significant for the simulation with
the smallest force softening length, but are under control for the
largest force softening length.
Thick line in the lower panels connects the average modulus of the
transverse velocities in bins of magnitude of radial velocity.
The visual impression gathered from the scatter plot is reinforced in
that with decreasing force softening length, we get larger transverse
motions. 

Figure~\ref{fig_plwave_atreepm} presents results of simulations with
the same initial conditions carried out with the Adaptive TreePM
code. 
We show the results for Adaptive TreePM without the $\gradeps$ term,
$n_n=16$ (left column); with the $\gradeps$ term, $n_n=16$ (middle
column), and, with the $\gradeps$ term, $n_n=32$ (middle column). 
In general we do not recommend use of $n_n=16$ due to reasons discussed
in the preceeding subsection on errors, discussion in the following
subsections, and in \citet{2007MNRAS.374.1347P}, but we nevertheless use it 
in order to look for early signs of two body collisions in a
simulation with a relatively small number of particles.
The phase portrait for all the three Adaptive TreePM runs is a faithful
representation of the expectations.   
The transverse motions are suppressed strongly, almost to the
same level as the TreePM simulation with a force softening of
$\epsilon=0.5$. 
One of the reasons for this is that the highest overdensities reached in this
experiment do not lead to a considerable reduction in the force softening
length.   
The other reason is that the adaptive nature of the code leads to presence of
a sufficient number of neighbours within a force softening length and hence
the anisotropy in the transverse force is reduced substantially.

Thick lines in the lower panels show the modulus of the
average transverse velocities in a few bins of the modulus of the
longitudinal velocity.  
We see that for TreePM, the magnitude of transverse motions drops
rapidly as we increase the force softening length. 
We also note that for the adaptive TreePM, the magnitude of transverse
motions is similar to that seen with the TreePM when $\epsilon$ for the
TreePM coincides with the $\epsmax$ for the ATreePM. 

We conclude the discussion of this test by noting that our requirement
of $\epsilon \gtrsim \bar{r}_{ij}$, where $\bar{r}_{ij}$ 
is the local inter-particle separation,
ensures collisionless behavior in evolution of the system.

\begin{figure*}
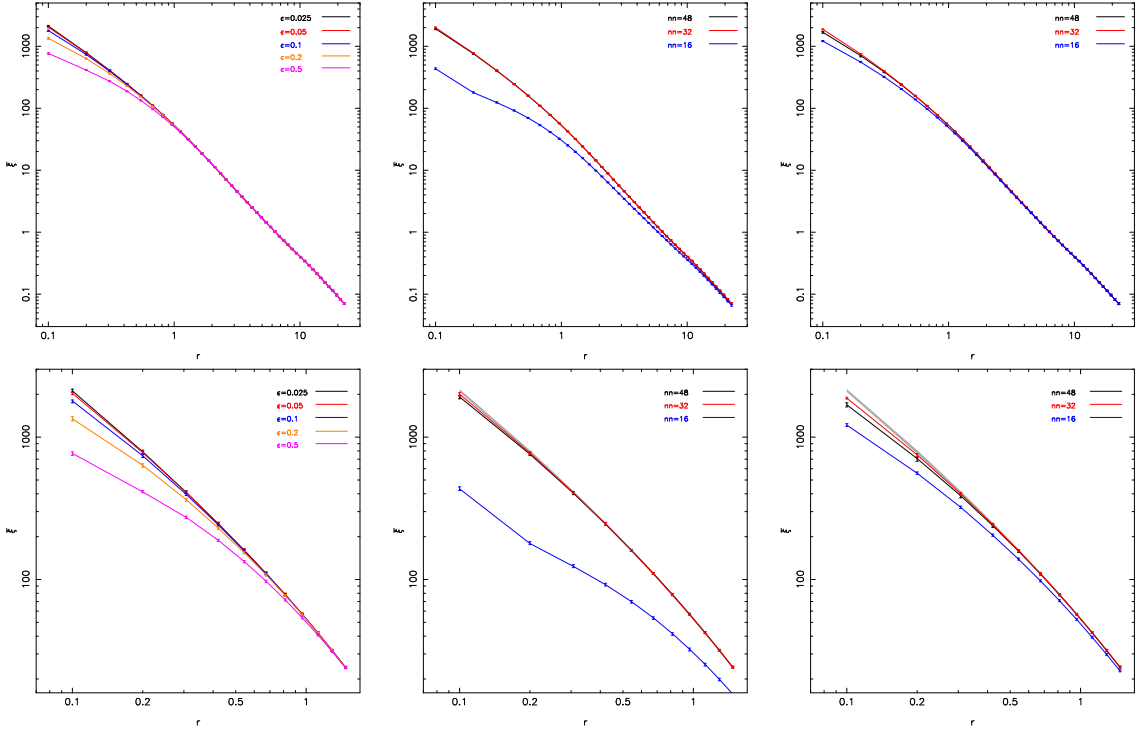

\begin{center}
\begin{tabular}{ccc}
\includegraphics[width=1.86truein]{5a.ps} &
\includegraphics[width=1.86truein]{5b.ps} &
\includegraphics[width=1.86truein]{5c.ps} \\
\includegraphics[width=1.86truein]{5d.ps} &
\includegraphics[width=1.86truein]{5e.ps} &
\includegraphics[width=1.86truein]{5f.ps} \\
\end{tabular}
\end{center}
\caption{{\emph{Top Row}}:
  This figure plots the volume averaged 2-point correlation
  function $\xibar$ as a function of scale at $\rnl = 6.0$ for TreePM 
  (left) and ATreePM with $\gradeps$ (middle) and ATreePM without
  $\gradeps$ (right). 
  The four curves (black, red, blue, ochre, mauve) for TreePM are for  
  runs with softening lengths of 
  $\epsilon = 
  \frac{r_s}{40},\frac{r_s}{20},\frac{r_s}{10},\frac{r_s}{5}, \frac{r_s}{2}$. 
  For the ATreePM runs the three curves (black, red, blue) are for 
  $n_n = 16, 32, 48$.
  {\emph{Bottom Row:}} zoomed in plots of Top Row. Grey surface in the
  ATreePM plots is $\xibar$ bounded by error-bars for TreePM 
  with $\epsilon=\frac{r_s}{40}$}
\label{fig_xibar_con}
\end{figure*}

\begin{figure*}
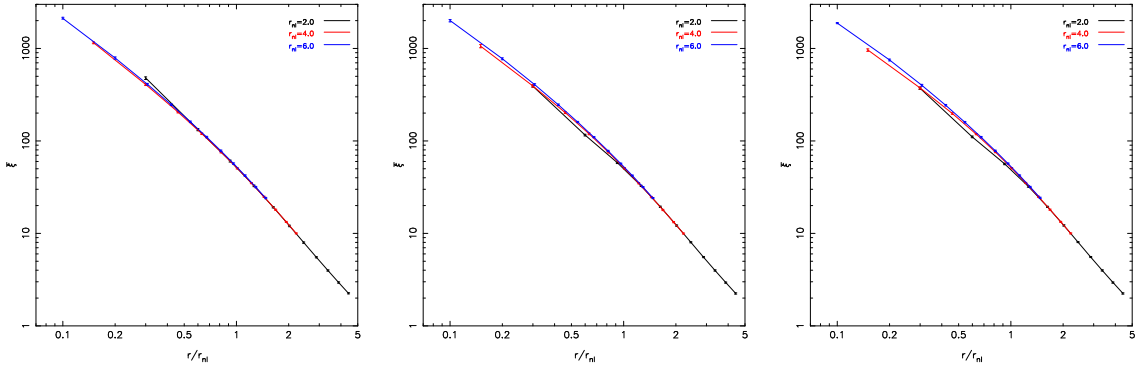

\begin{center}
\begin{tabular}{ccc}
\includegraphics[width=1.86truein]{6a.ps} &
\includegraphics[width=1.86truein]{6b.ps} &
\includegraphics[width=1.86truein]{6c.ps} \\
\end{tabular}
\end{center}
\caption{This figure tests self similar evolution of the scaled 2-point 
correlation function $\xibar(r/r_{nl})$ 
by plotting it at the epochs when 
$r_{nl}=2.0,4.0,6.0$ (black, red, blue lines)for TreePM (left) ATreePM
with $\gradeps$ (middle) , ATreePM without $\gradeps$}
\label{fig_xibar_ss}
\end{figure*}

\subsection{Convergence with $n_n$ and relevance of the $\gradeps$ term}

For the following discussions we run a power-law model with index $n=-1.0$. 
The spectrum is normalized at a an epoch when the scale of non-linearity $\rnl
= 6.0$ in grid units.  
The scale of non-linearity is defined as the scale at which the linearly
extrapolated mass variance, defined using a top hat filter, is unity
$\sigma(\rnl,z)=1$.  
The simulations are done with $128^3$ particles and a $128^3$ grid.
We choose $r_s=1$ grid length.  
Other simulation parameters are as described in the subsection \S{3.1}
on the TreePM code.
We assume that the Einstein-de Sitter model describes the background
universe.
In this case self-similar evolution of quantities for power law models
provides an additional test for simulation results. 

\subsubsection{Clustering Properties}

We compute the volume averaged $2-$point correlation function
$\xibar$.
\beq
\xibar(r) = \frac{3}{r^3}\int^r_0 \xi (x)x^2 dx
\eeq
$\xi$ is the two point correlation function 
\citep{1980lssu.book.....P}.
To compute $\xibar$ from simulation output, we take $15$ independent
random subsets of $10^5$ particles each. 
We then estimate the average value of $\xibar$ over these subsets. 
The maximum and the minimum values of $\xibar$ in these subsets are
our estimate of the errors.
Figure~\ref{fig_xibar_con} shows $\xibar$ at an epoch when 
$\rnl = 6.0$.
TreePM (left column) with $\epsilon = 0.025$, $0.05$, $0.1$, $0.2$, $0.5$ are
shown in black, red, blue, ochre and magenta respectively. 
ATreePM with $\gradeps$ (middle column) and ATreePM without $\gradeps$ (right
column) with $n_n=16$, $32$, $48$ are shown in blue, red, black lines
respectively.  
The lower row is a zoom-in of $\xibar$ so as to highlight
the differences at small scales, due to the variation in $\epsilon$
and $n_n$ in different runs. 

One can show that we require a minimum of $12$ neighbors to solve for
Eqn.(\ref{eps_con}).  
The $\gradeps$ term is important in over-dense regions where $\epsilon_i \ll
\epsmax$, and is a rapidly varying function of density, one therefore requires
a reasonably large $n_n$ to compute it accurately. 
Use of a small $n_n$ leads to a noisy estimate of $\gradeps$ and
Figure~\ref{fig_xibar_con} illustrates this point\footnote{Also see
  Figure~\ref{fig_err}.}.   
With $n_n=16$, $\xibar$ deviates significantly from our expectations. 
Disagreement is worse for ATreePM with $\gradeps$, as this is the case
where a poor estimate of the extra term leads to larger errors.
If we use a larger $n_n$, we expect to see convergence at some point. 
Curves for $n_n=32$ and $48$ agree with each other (within error-bars) 
for ATreePM with $\gradeps$ indicating that evaluation of the extra 
term is stable.  
It also indicates that this extra term compensates for the use of a
larger number of particles, otherwise typical softening length is
expected to be larger for higher $n_n$ and this should lead to
lowering of the clustering amplitude at small scales.
This is very clearly illustrated for TreePM 
(left column in Figure~\ref{fig_xibar_con}) where increasing  
$\epsilon$  reduces $\xibar$ monotonically.
We also see this effect for ATreePM without the $\gradeps$ term where
the amplitude of clustering is much smaller for $n_n=48$ as compared
to $n_n=32$ and the difference between the two is more than twice that
seen for ATreePM with the $\gradeps$ term.
Such a behavior is expected, and was also seen by \cpricet. 
They noted that the increase in force softening length with $n_n$
leads to a bias in the force at small scales. 
The $\gradeps$ term corrects for this and biasing of force 
is less important. 
Since the distribution of particles gets more strongly clustered with time, 
we expect over-softening of the force field to degrade further
evolution for the TreePM and the ATreePM without $\gradeps$, beyond a
certain epoch. 

We have plotted the scaled 2-point volume averaged correlation
$\xibar(r/r_{nl})$ at small scales in Figure~\ref{fig_xibar_ss} for TreePM
with $\epsilon=0.025$ (left panel), ATreePM with the $\gradeps$ term (middle
panel) and ATreePM without the $\gradeps$ term (right panel). 
We used simulations with $n_n=32$ for both the ATreePM runs shown here. 
$\xibar$ has been computed at epochs when the scale of non-linearity
$\rnl=2$, $4$ and $6$ (black, red and blue lines respectively). 
Since the only scale in power-law models is the scale of non-linearity $\rnl$,
introduced by gravity, one expects $\xibar$ to evolve in a self-similar manner.
The scale of self-similarity, $\rss$, for a given epoch is is determined by
finding the scale at which $\xibar$ matches with the $\xibar$ of the earlier
epoch within the error bars computed in the manner described above.

\begin{table*}
\caption{The following table shows the variation of $\rcon$ and $\rss$
with time for TreePM and the two variants of ATreePM 
as a function of softening} 
\vspace{0.3cm}
\begin{center}
\begin{tabular}{||l|l|l|l|l|l||} 
\hline\hline
Runs 
& $\rcon$  
& $\rcon$
& $\rcon$
& $\rss$
& $\rss$
\\ 
& $\rnl = 6.0 $
& $\rnl = 4.0 $
& $\rnl = 2.0 $
& $\rnl = 6.0 $
& $\rnl = 4.0 $\\
\hline\hline
TreePM ($\epsilon = \frac{r_s}{2}$) & 1.2 & 0.96 &  0.61 & 1.1 & 0.67\\ \hline
TreePM ($\epsilon = \frac{r_s}{5}$) & 0.61 & 0.44 & 0.36 & 0.42 & 0.15\\ \hline
TreePM ($\epsilon = \frac{r_s}{10}$) & 0.25 & 0.20 &  0.10 & 0.18 & 0.16\\ \hline
TreePM ($\epsilon = \frac{r_s}{20}$) & 0.10 & 0.10 & 0.10 & 0.10 & 0.17\\ \hline
TreePM ($\epsilon = \frac{r_s}{40}$) & \,\,- & \,\,- &  \,\,- & 0.10 & 0.18\\ \hline
ATreePM ($\gradeps, n_n=32$) & 0.10 & 0.15 &  0.30 & 0.27 & 0.31\\ \hline
ATreePM ($\gradeps, n_n=48$) & \,\,- & \,\,- & \,\,-  & 0.42 & 0.47\\ \hline
ATreePM ($n_n=32$)& 0.27 & 0.28 &  0.38 & 0.54 & 0.42\\\hline
ATreePM ($n_n=48$)& \,\,- & \,\,- &  \,\,- & 0.59 & 0.50\\
\hline 
\hline
\end{tabular}
\end{center}
\vspace{0.3cm}
\label{tab_xibar_con}
\end{table*}

We illustrate numerical convergence properties of the three codes in
table~\ref{tab_xibar_con}. 
Here, we list the scale $\rcon$ and $\rss$ at different epochs.
For ATreePM $\rcon$ is the scale beyond which $\xibar$ with the given
$n_n$ matches with a reference $\xibar$ with $n_n=48$. 
For TreePM this is the scale where $\xibar$ with a given force
softening length matches with the reference $\xibar$ with
$\epsilon=0.025$. 

We find that both variants of ATreePM converge with time, 
the convergence being more rapid for ATreePM with $\gradeps$ as
compared to ATreePM without $\gradeps$. 
The scale of convergence $\rcon$ is also smaller for ATreePM with
$\gradeps$. 
On the other hand self-similar behavior is degraded with evolution
for ATreePM without $\gradeps$ as $\rss$ increases with time, whereas
for ATreePM with the $\gradeps$ term the evolution is self-similar
over a larger range of scales.

Thus we may conclude that the ATreePM with the $\gradeps$ has well
defined numerical convergence as we vary $n_n$, and the evolution of
power law models for this code is self-similar over a wide range of
scales.  
This sets it apart from the ATreePM without the $\gradeps$ term, where
the convergence with $n_n$ is not well defined and the evolution of a
power law model is self-similar over a smaller range of scales. 
Better match with the fixed resolution TreePM code is an added
positive feature for the code with the $\gradeps$ term.
This raises obvious questions about AMR codes, where no such term is
taken into account as we increase the resolution of the code.
We comment on this issue in the Discussion section.

We briefly comment on the convergence and self-similar evolution in
the fixed resolution TreePM code.
As seen in Table~\ref{tab_xibar_con}, we find that at early times the
scale above which we have self-similar evolution is almost the same
for $\epsilon \leq 0.2$. 
This may indicate discreteness noise, or it may be due to two body
collisions during early collapse
\citep{1998ApJ...497...38S,1997ApJ...479L..79M,2008arXiv0805.1357J,2008ApJ...686....1R}. 
At late times, the scale above which evolution is self-similar is
$\rss \sim 2\epsilon$.  
This indicates that any transients introduced at early times by
discreteness, etc. have been washed out by the transfer of power from
large scales to small scales 
\citep{1991MNRAS.253..295L,1992ApJ...394L...1E,1997MNRAS.286.1023B,2008arXiv0802.2796B}.
Thus one gets an improved self-similar evolution at late times in the
fixed resolution TreePM with the use of a smaller softening length.

On the other hand we see that for TreePM, $\rcon$ increases with
time. 
As defined above, this is the scale at which $\xibar$ obtained with a given
value of $\epsilon$ matches with the value obtained using $\epsilon=0.025$.  
We find that at the last epoch $\rcon \geq 2 \epsilon$, whereas at
early times the limit is $\rcon \geq \epsilon$.
This trend is contrary to that seen with the ATreePM for which $\rcon$
decreases with time, the rate at which it comes down being faster for
the ATreePM with the $\gradeps$. 
We would like to point out that we do not probe $\xibar$ for scales $r
< 0.1$ as the number of pairs with a smaller separation is often too
small to get reliable estimates of $\xibar$.

Another remarkable fact is that for the TreePM code, $\rss \leq
\rcon$.   
Thus we have the desired behavior in terms of evolution even though
we are yet to achieve numerical convergence at the relevant scales. 
Such a problem does not exist for the ATreePM code.

Lastly, as we show below, the ATreePM code is very good at resolving highly
over-dense regions very well.  
This would appear to be in contradiction with the slightly lower two point
correlation function. 
The reason for the slightly lower correlation function is that the ATreePM
code does not resolve small haloes, in particular those with fewer than $n_n$
particles (see the discussion of mass function of collapsed haloes).
Indeed, the number density of small haloes is severely under-estimated in the
ATreePM simulations and we believe that this is the main reason for a weaker
two point correlation function.

\subsubsection{Collapsed Haloes}

\begin{figure*}
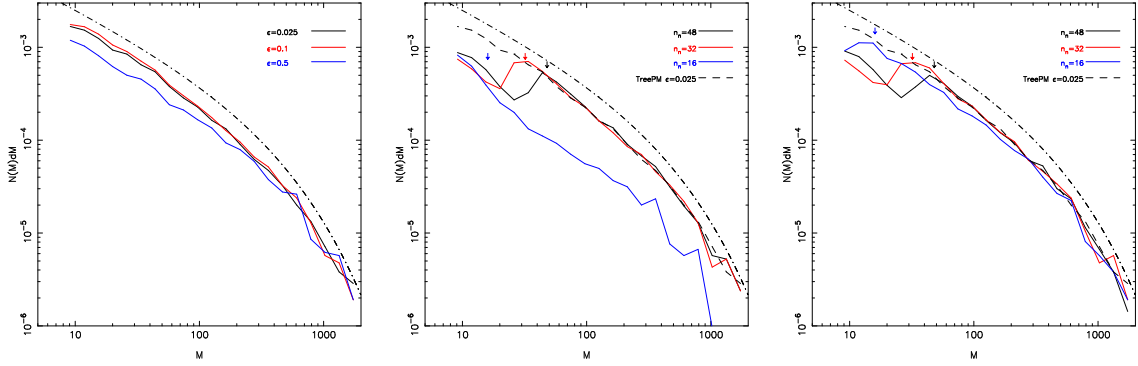

\begin{center}
\begin{tabular}{ccc}
\includegraphics[width=1.86truein]{7a.ps} &
\includegraphics[width=1.86truein]{7b.ps} &
\includegraphics[width=1.86truein]{7c.ps} \\
\end{tabular}
\end{center}
\caption{Mass function for TreePM (left), ATreePM with $\gradeps$ (middle)
and ATreePM without $\gradeps$ (right). For TreePM $\epsilon =
\frac{r_s}{40}, \frac{r_s}{10}, \frac{r_s}{2}$ are drawn in black,
red, blue lines respectively. $n_n= 32,48$ are drawn in black and red
lines for ATreePM. The arrows represent $n_n$ in mass units. The black
dot-dashed curve on every plot is the Press-Schecter mass function,
shifted vertically by multiplying by a factor of 1.5. The black dashed
line in ATreePM is for TreePM with $\epsilon=\frac{r_s}{40}$}
\label{fig_massfn}
\end{figure*}
\begin{figure*}
\begin{center}
\begin{tabular}{cc}
\includegraphics[width=2.8truein]{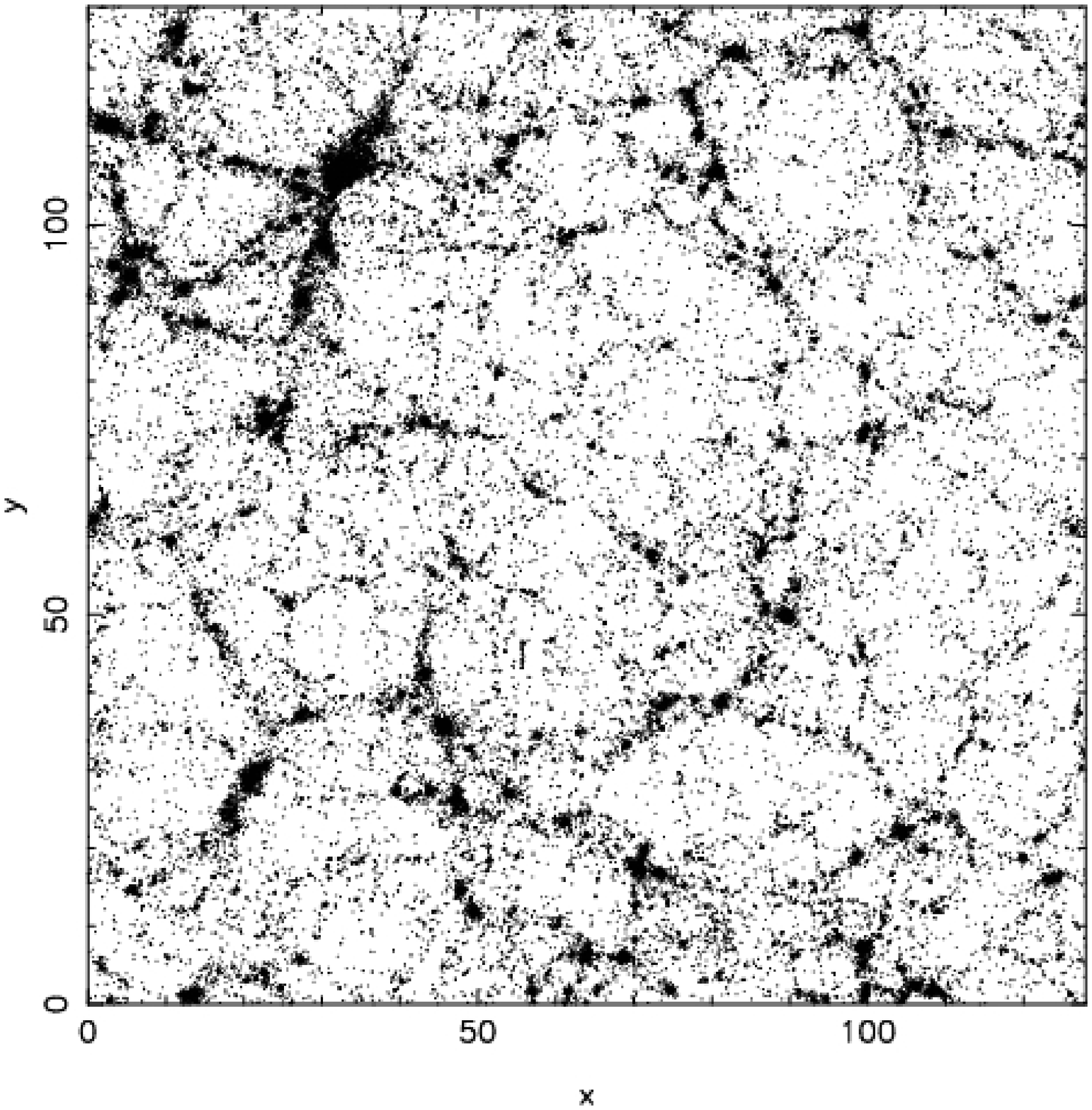} &
\includegraphics[width=2.8truein]{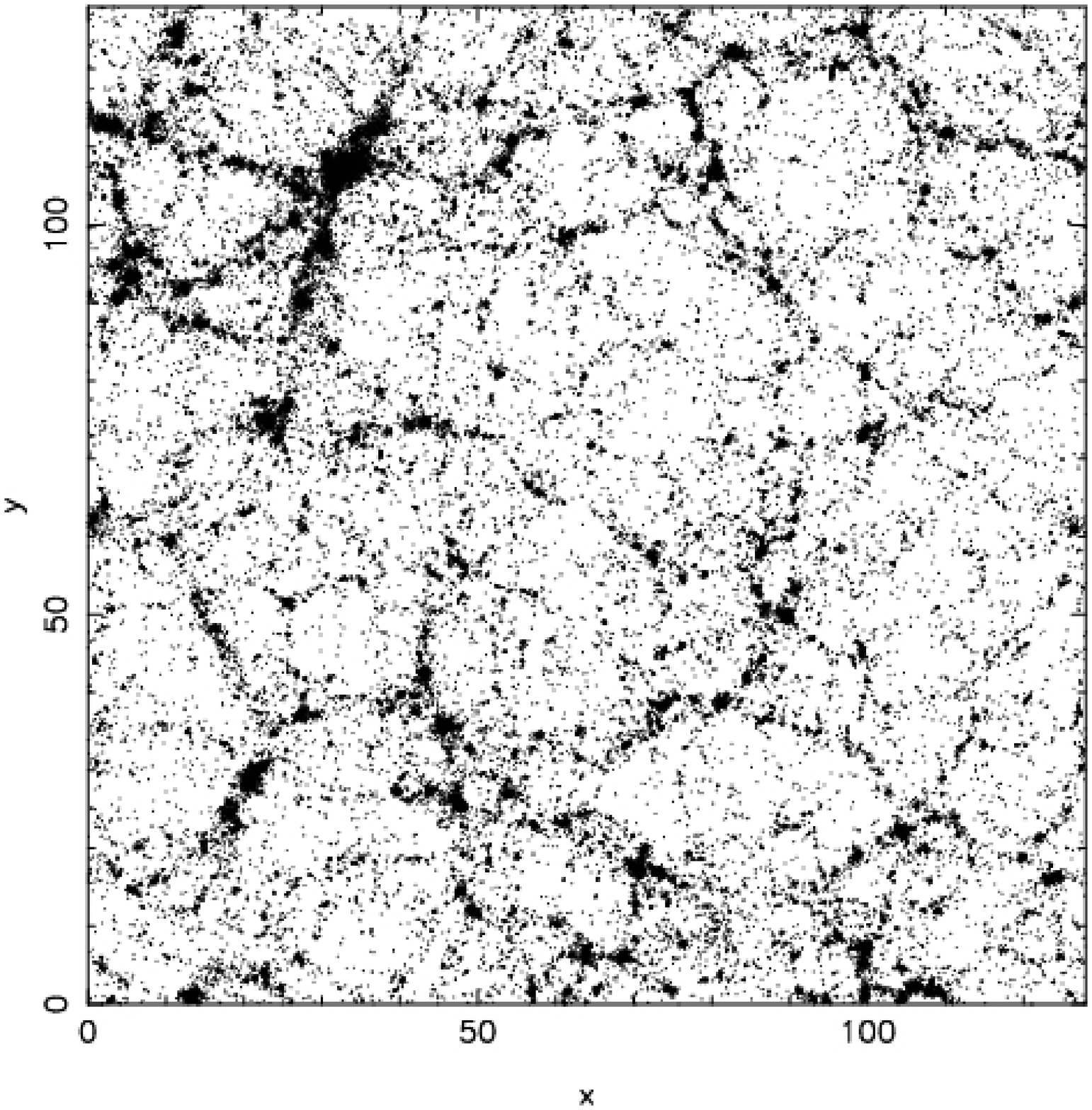} \\
\includegraphics[width=2.8truein]{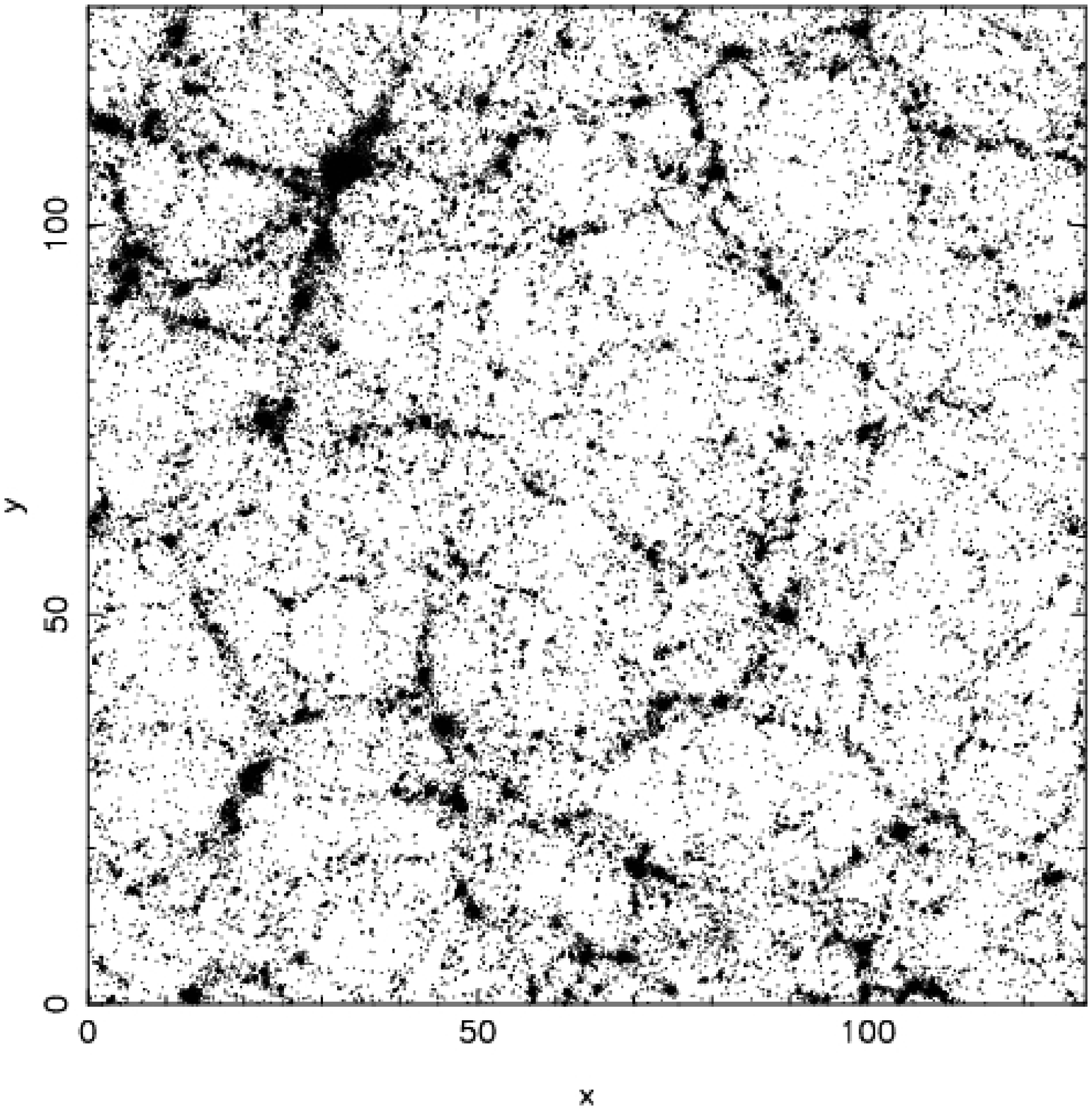} &
\includegraphics[width=2.8truein]{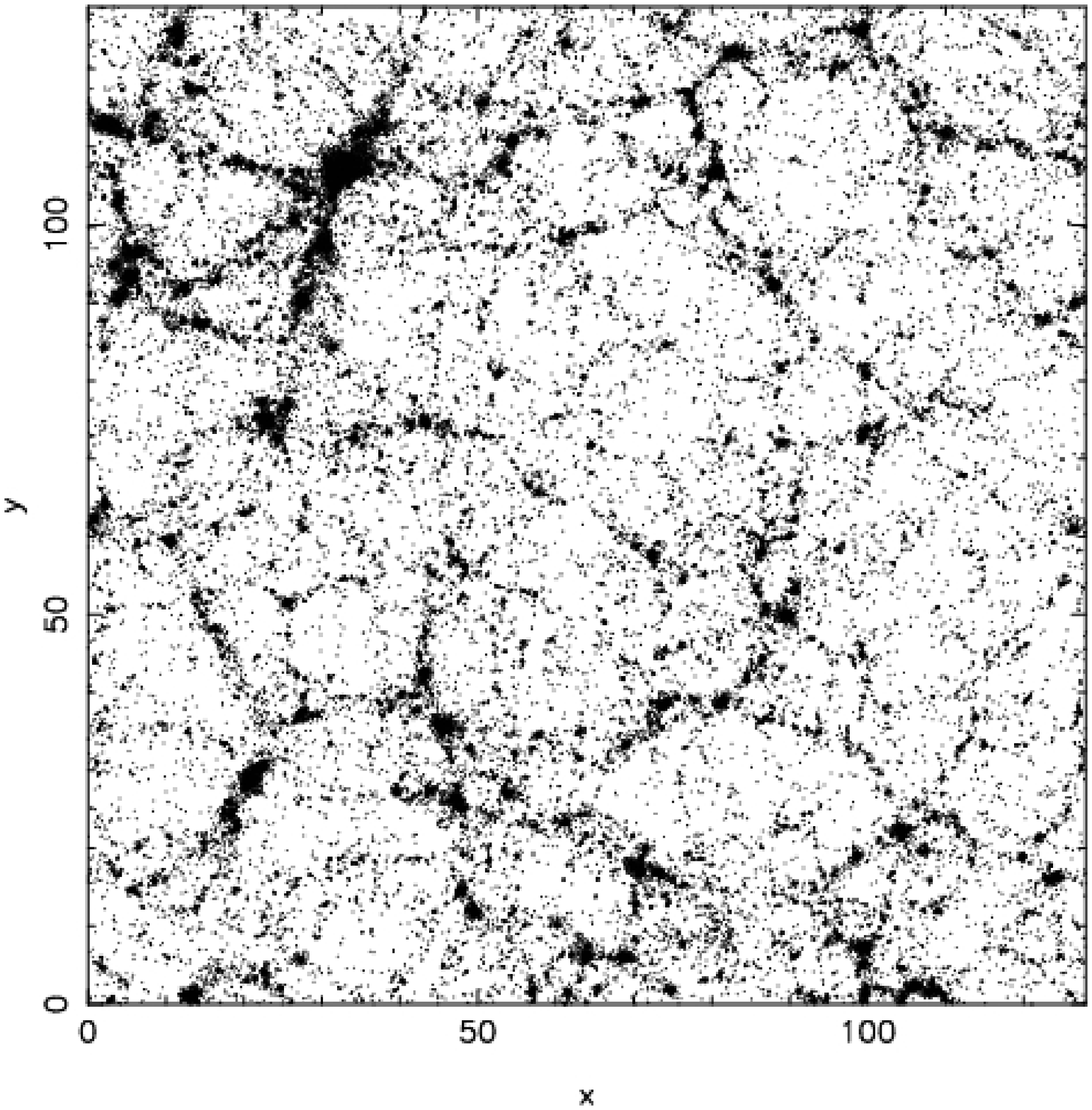} \\
\end{tabular}
\end{center}
\caption{This figure shows slice from simulation where the most
  massive halo resides (see top left corner). Top panel is for TreePM 
  with $\epsilon=\frac{r_s}{40}$ (left) and ATreePM with  
  $\epsilon=\frac{r_s}{2}$. Bottom panel is for ATreePM 
  with $\gradeps$ (left) and ATreePM without $\gradeps$ (right).
  $n_n=32$ was taken for ATreePM.
}
\label{fig_slice_full}
\end{figure*}

\begin{figure*}
\begin{center}
\begin{tabular}{cc}
\includegraphics[width=2.8truein]{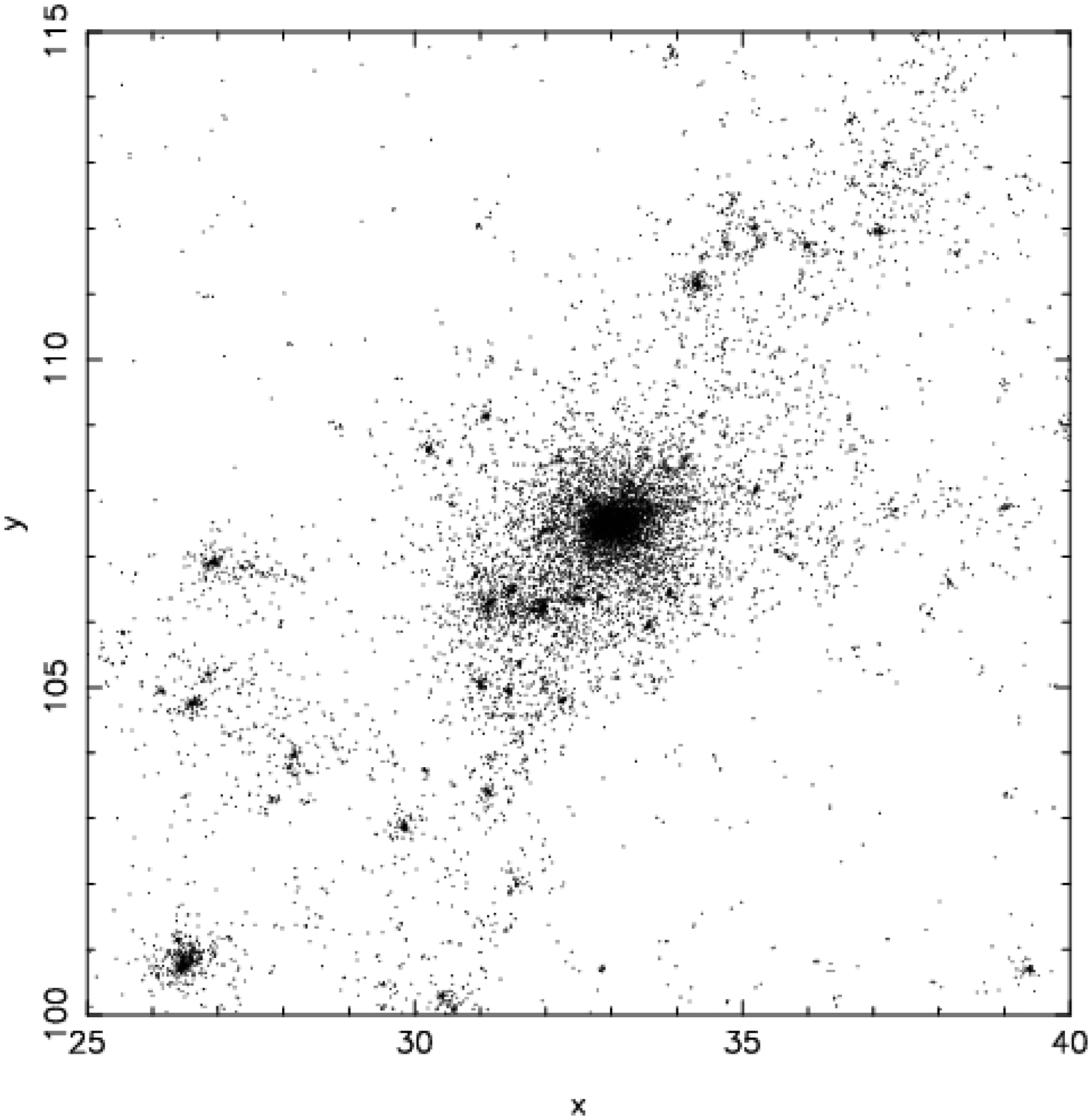} &
\includegraphics[width=2.8truein]{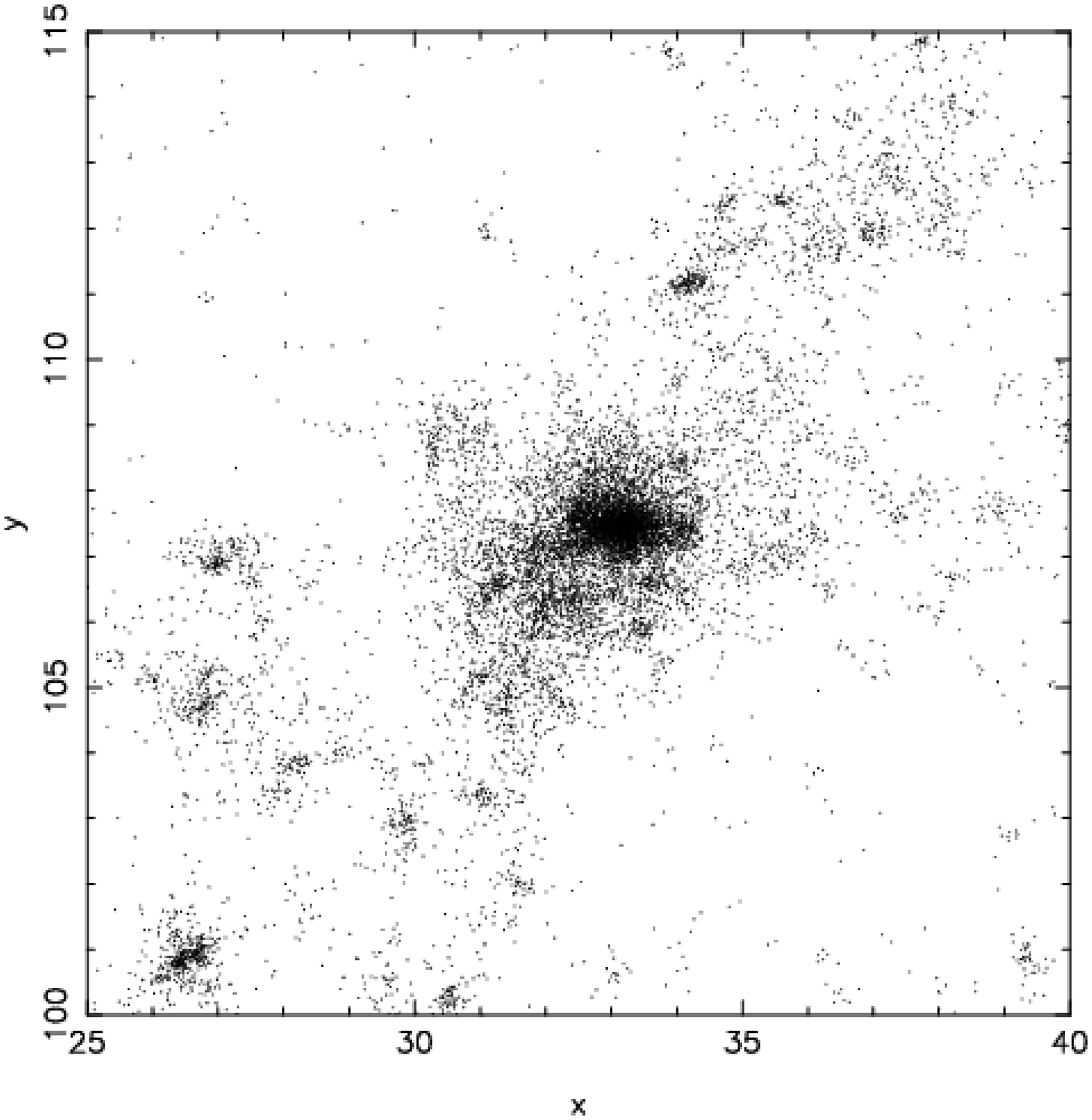} \\
\includegraphics[width=2.8truein]{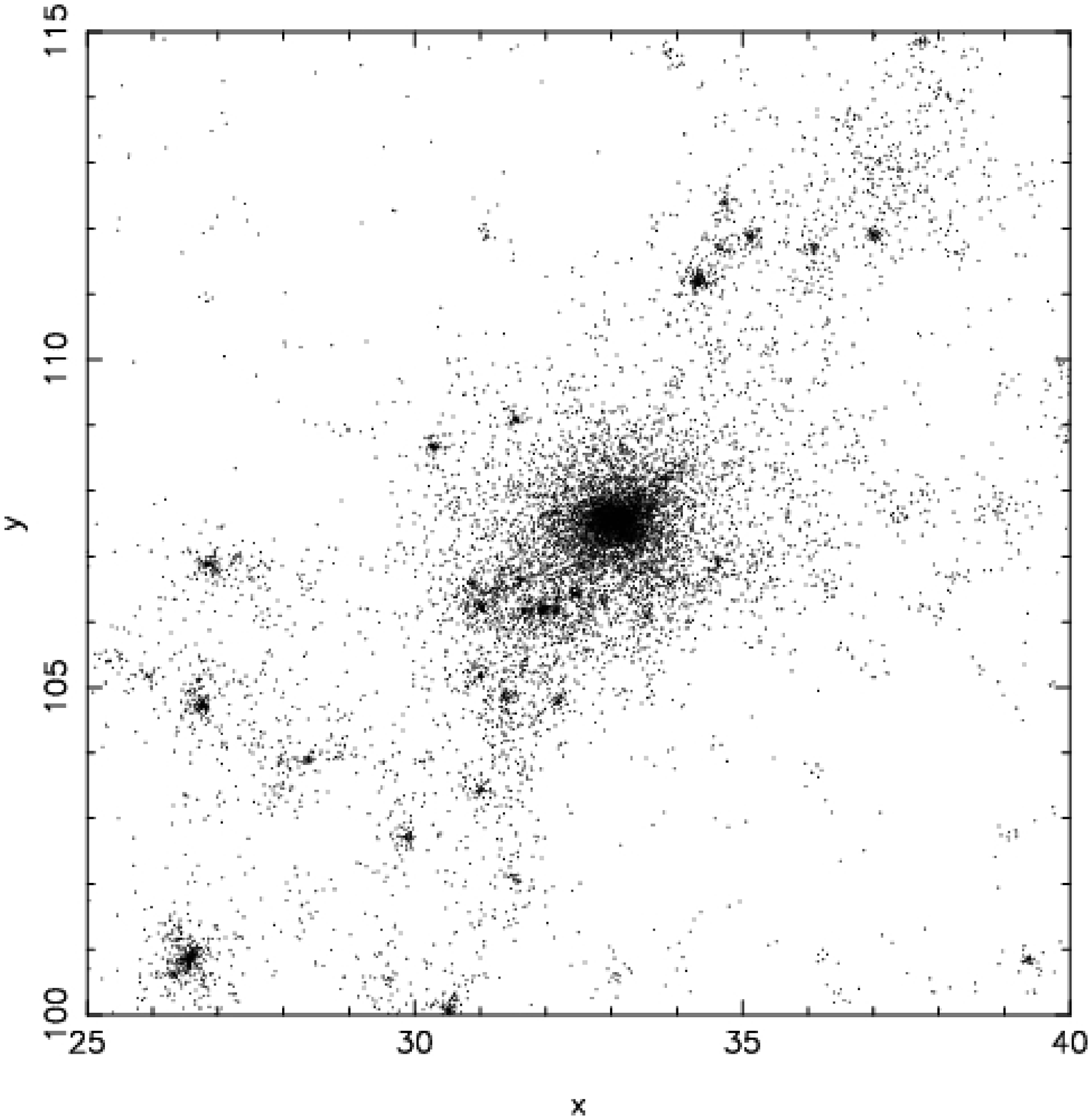} &
\includegraphics[width=2.8truein]{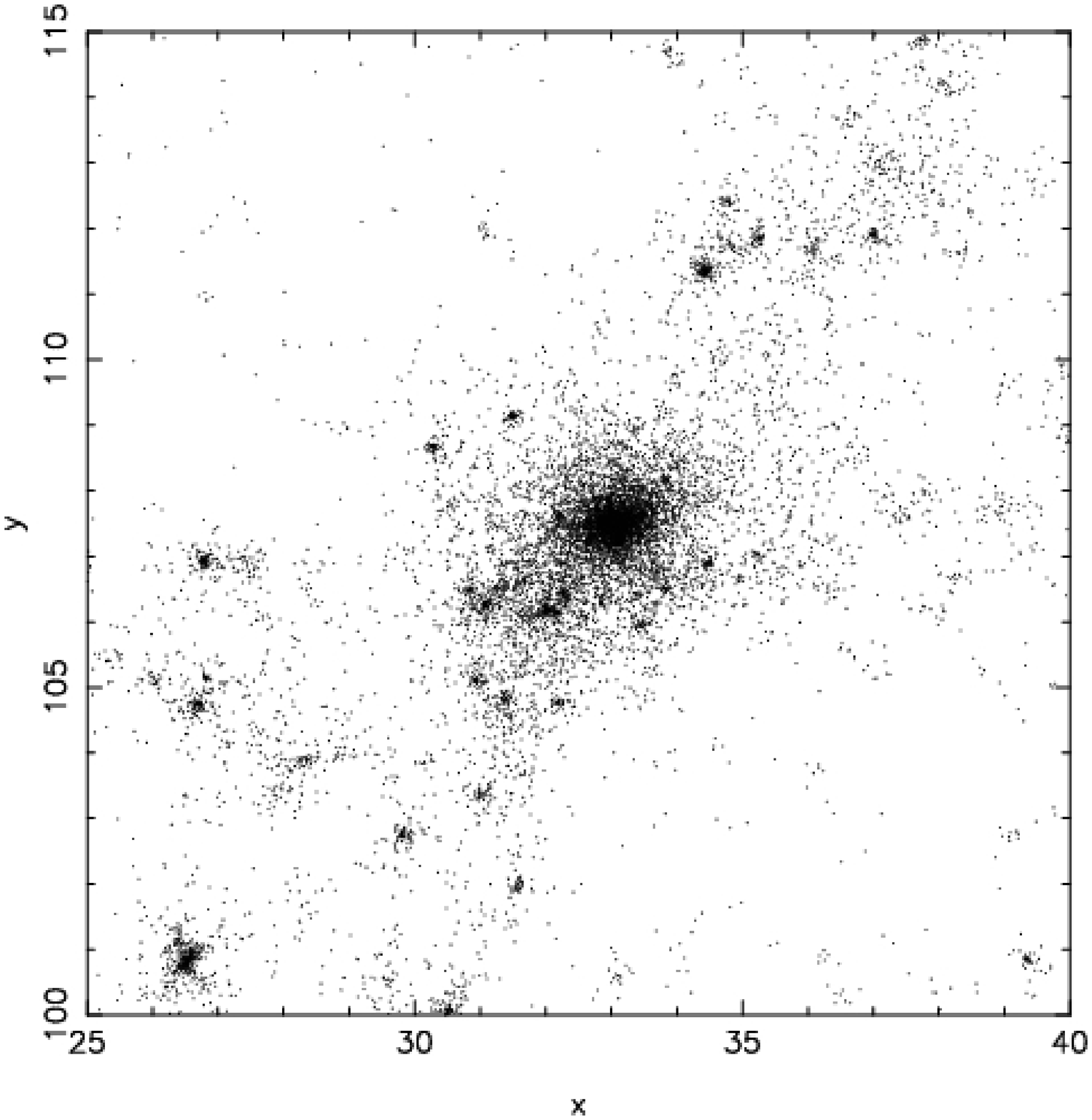} \\
\end{tabular}
\end{center}
\caption{This figure is a zoom in of Fig.~\ref{fig_slice_full}
  and shows the region of simulation where the most
  massive halo resides (see top left corner of
  Fig.~\ref{fig_slice_full}). 
  Top panel is for TreePM 
  with $\epsilon=\frac{r_s}{40}$ (left) and ATreePM with  
  $\epsilon=\frac{r_s}{2}$. Bottom panel is for ATreePM 
  with $\gradeps$ (left) and ATreePM without $\gradeps$ (right).
  $n_n=32$ was taken for ATreePM.
}
\label{fig_slice_zoom}
\end{figure*}
\begin{figure*}
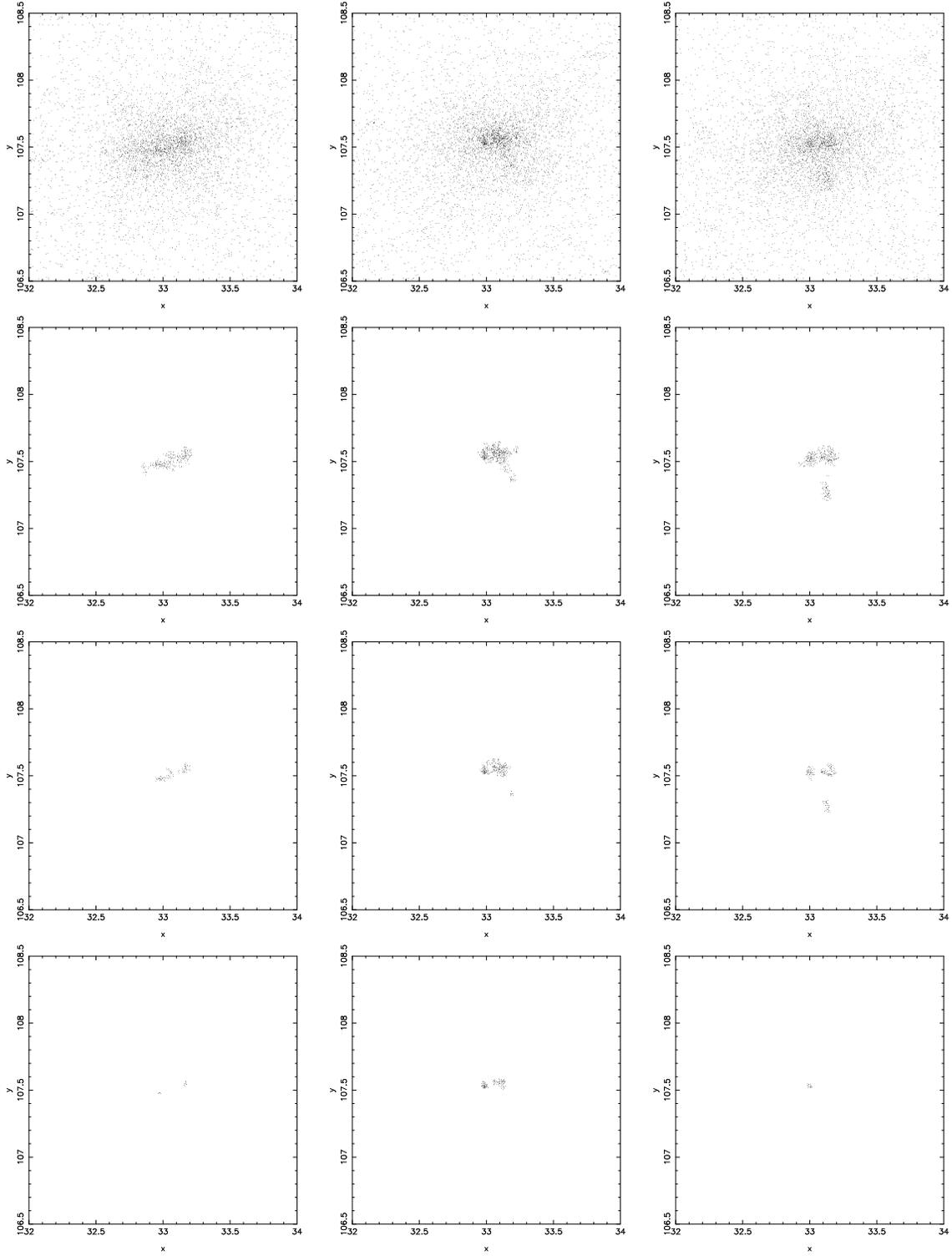

\begin{center}
\begin{tabular}{ccc}
\includegraphics[width=1.86truein]{10z1a.ps} &
\includegraphics[width=1.86truein]{10z1b.ps} &
\includegraphics[width=1.86truein]{10z1c.ps} \\
\includegraphics[width=1.86truein]{10z2a.ps} &
\includegraphics[width=1.86truein]{10z2b.ps} &
\includegraphics[width=1.86truein]{10z2c.ps} \\
\includegraphics[width=1.86truein]{10z3a.ps} &
\includegraphics[width=1.86truein]{10z3b.ps} &
\includegraphics[width=1.86truein]{10z3c.ps} \\
\includegraphics[width=1.86truein]{10z4a.ps} &
\includegraphics[width=1.86truein]{10z4b.ps} &
\includegraphics[width=1.86truein]{10z4c.ps} \\
\end{tabular}
\end{center}
\caption{This figure shows the largest halo at $\rnl=6.0$ for TreePM
  with $\epsilon=\frac{r_s}{40}$ (left column), ATreePM with $\gradeps$ and
  $n_n=32$ (middle column) and ATreePM without $\gradeps$ and
  $n_n=32$ (right column). The first row contains all the particles
  in the halo. The second, third and forth rows represent 
  particles in the halo which have a density greater than
  $5\times10^4$, $10^5$ and $2\times10^5$ respectively. 
}
\label{fig_halo}
\end{figure*}

We now look at another global indicator, namely the mass function of
collapsed halos, to compare the performance of the TreePM and the
ATreePM codes. 
The number density of halos $N(M)dM$ in the mass range $(M,M+dM)$ is plotted
in Figure~\ref{fig_massfn}.   
Halos were identified using the Friends-Of-Friends algorithm with a linking
length $l_l=0.1$ in grid units and only haloes with at least $8$ particles
were considered. 

For TreePM we see that as we increase the force softening length there
is a clear change in the mass function.  
The mass function converges for all masses with $\epsilon \leq 0.2$. 
However, the mass function for $\epsilon=0.5$ shows fewer haloes of up
to about $10^2$ particles. 
Thus the effect of a large softening length is seen in mass function
up to fairly large mass scales.
This qualitative behavior is insensitive to the choice of linking
length, though the scale of convergence is smaller if we choose a
larger linking length. 

The ATreePM does not sample the haloes with fewer than $n_n$ members
properly. 
This is expected as such over-densities are not sufficient to reduce
the softening length from $\epsmax$. 
At larger scales we have a convergence for $n_n=32$ and $n_n=48$,
where the mass function agrees with that for TreePM with $\epsilon=0.025$. 
Mass function for $n_n=16$ deviates significantly from the other two for
ATreePM, deviation being larger for the ATreePM with the $\gradeps$ term. 
We believe that this is due to the noisy estimation of the force softening
length and the $\gradeps$ term, and resulting errors in force.

As a reference the Press-Schecter mass function is also 
plotted in all three panels (black dot-dashed line). 
However it has been shifted up vertically by multiplying $N(M)dM$ by a 
factor of $1.5$ for ease of comparison.

Distribution of particles in a simulation is a useful representation for
comparing large scale features.
The distribution of particles in a thin slice is shown in 
Figure~\ref{fig_slice_full}.  
We have shown the distribution for four simulations: TreePM with
$\epsilon=0.025$ (top-left), TreePM with $\epsilon=0.5$ (top-right), ATreePM
with the $\gradeps$ term and $n_n=32$ (lower-left) and ATreePM without the
$\gradeps$ term and $n_n=32$ (lower-right). 
The distribution of particles is for the last epoch, corresponding to
$r_{nl}=6$ grid lengths. 
This slice contains the largest halo in our simulation on the top left corner.
We note that the large scale distribution is the same in all cases, indicating
that the Adaptive TreePM is not changing anything at large scales.

We zoom in on the region around the largest halo in 
Figure~\ref{fig_slice_zoom}, where the
distribution of particles is shown from the set of simulations used in
Figure~\ref{fig_slice_full}. 
Distribution of mass at scales larger than a gid length is the same in all
simulations but at small scales we begin to see some differences between the
distribution of particles in different simulations. 
TreePM with $\epsilon=0.5$ differs most from the other three, in that it does
not resolve small scale structures. 
This happens due to the relatively large force softening length that inhibits
collapse if the expected size of the collapsed halo is smaller than
$\epsilon$. 
The notable differences between the TreePM ($\epsilon=0.025$) and the ATreePM
slices are as follows:
\begin{itemize}
\item
Small haloes in the region away from the large haloes are more compact for
TreePM than for ATreePM.  
This is perhaps caused by the ATreePM not having a constant $\epsilon$ and it
is likely that in these clumps the value of $\epsilon$ is larger than $0.025$
that is used in the TreePM. 
\item
Location of sub-structure in the central halo differs somewhat between the
different simulations.
\end{itemize}

Figure~\ref{fig_halo} shows the inner regions of the halo seen in 
Figure~\ref{fig_slice_zoom}. 
We show the inner parts of the halo in the four cases (TreePM
($\epsilon=0.025$), ATreePM with the $\gradeps$ term
($n_n=32$), ATreePM without the $\gradeps$ term ($n_n=32$). 
The top row shows all the particles in the inner parts of the halo. 
The second row shows all the particles with an SPH over-density estimate
(computed with $n_n=32$ and using the kernel specified in
Eqn.~\ref{eq_mon_kernel})
of greater than $5 \times 10^4$, the third row shows the particles
with an over-density of greater than $10^5$, and, the last row shows particles
with an overdensity larger than $2 \times 10^5$.  
We see that the ATreePM with the $\gradeps$ term shows the most pronounced
central part of the halo in the top panel, and this impression gains strength
as we move to lower panels. 
Table~\ref{table_halo} gives us the number of particles 
in the various panels of Fig.~\ref{fig_halo}. 
TreePM with $\epsilon=0.025$ manages to trace some of the highly over-dense
substructure seen in the ATreePM simulations, though with a much smaller
number of particles in these clumps. 
The ATreePM with the $\gradeps$ term does better than the one without,
particularly at the highest over-densities used here. 
Indeed  ATreePM with the $\gradeps$ term retains nearly $7$ times 
as many particles in the halo core as compared to TreePM and 
ATreePM without the $\gradeps$ term.
The same thing is also reflected in Figure~\ref{fig_xibar_con} 
where the ATreePM without the $\gradeps$ term is seen to underestimate
clustering at small scales. 
It is noteworthy that the size of highly over-dense structures in the core of
this halo are much bigger than the force softening length for the TreePM.

It is notable that the ATreePM is able to resolve highly over-dense regions
while taking much less time than the fixed resolution TreePM.  
Thus we have the added performance at the cost of fewer resources. 

\begin{table}
\caption{This table shows the number of particles $n_{part}$
  above a density threshold (as in Figure~\ref{fig_halo})
  retained in the core of the most massive halo 
}
\vspace{0.1cm}
\begin{center}
\begin{tabular}{||l|l|l|l||} 
\hline\hline
Runs 
& $n_{part}$   
& $n_{part}$ 
& $n_{part}$ 
\\ 
& $\rho\geq 5\times 10^4$
& $\rho\geq 10^5$
& $\rho\geq 2\times 10^5$
\\
\hline 
TreePM ($\epsilon = \frac{r_s}{40}$) &320 &106 &18
\\
\hline 
ATreePM ($\gradeps, n_n=32$) &568 &297 &125 
\\
\hline 
ATreePM ($n_n=32$) &374 &169 &18 
\\
\hline 
\hline
\end{tabular}
\end{center}
\label{table_halo}
\end{table}

\subsubsection{Dynamics within Collapsed Haloes}

The equation of motion for the adaptive code has the additional $\gradeps$
term, 
and it is important to test whether this term leads to a change in dynamics in
highly overdense regions. 
We test this by studying the dynamics in central cores of the largest haloes
in the simulation. 
We study the ratio $\langle 2T \rangle / \langle U \rangle$ at a scale
corresponding to $r_{200}/5$, with the averaging done around the centre of
mass of the halo. 
We chose the reference scale in this manner as the density profiles for haloes
with different softening length differ considerably and using a high density
contour to define the radius leads to very different physical scales. 
Figure~11 shows $\langle 2T \rangle / \langle U \rangle$ as a function of
softening length $\epsilon$ for the fixed resolution TreePM simulations. 
This has been done for the five largest haloes in the simulation. 
We see that as we go to a larger softening length, the ratio becomes larger
than unity, indeed for $\epsilon=0.5$ the ratio is closer to two for one of
the haloes.  
This is likely to happen if the size of the over-dense core is comparable to
the softening length, and indeed this is the case for some of the haloes used
for this plot. 
The same plot shows the ratio as seen in the adaptive code (with the $\gradeps$
term) with empty squares, and as filled triangles for the adaptive code without
the $\gradeps$ term. 
These circles are plotted at small values of $\epsilon$, not used for TreePM
runs. 
The values of $\epsilon$ used for positioning these symbols in the plot has
no relevance to the simulation, and this has been done purely for the purpose
of plotting the values. 
We see that the values for the ratio in both the adaptive codes correspond
closely to the values of the ratio seen in fixed resolution TreePM simulations
with a small $\epsilon$. 
Thus we may conclude that the additional term in the Adaptive code is not
leading to any significant changes in the dynamics, as compared to the fixed
resolution codes. 

\begin{figure}
\begin{center}
\begin{tabular}{c}
\includegraphics[width=2.8truein]{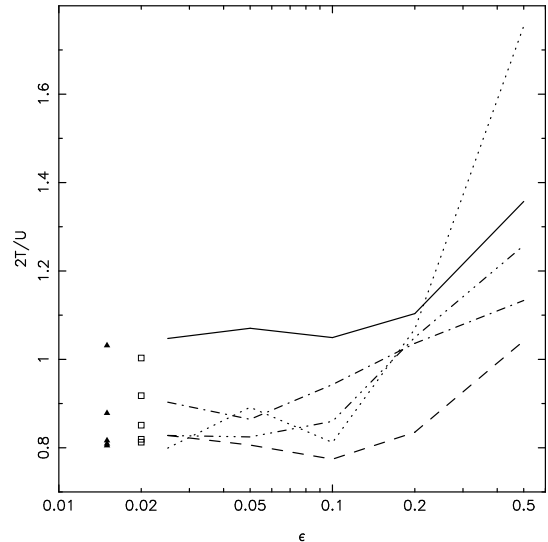} 
\end{tabular}
\end{center}
\caption{Ratio of two times the kinetic energy and the potential energy,
  $2T/U$, for the 5 largest halos as a function of softening length for 
TreePM (black lines). Open squares and filled triangles are for ATreePM 
with and without $\gradeps$ respectively.}
\label{fig_2tu}
\end{figure}

\section{Discussion}

In this paper we have introduced the first cosmological N-Body code that has a
continuously varying force resolution. 
This is based on a well defined formalism that ensures energy and momentum
conservation. 
An important aspect of this formalism is that it modifies the equation of
motion at scales below the force softening length.  
Our implementation of the formalism uses methods developed for SPH codes, this
is required as we need to assign quantities like density to particles. 
We have described our implementation of the code in detail here. 
We have given a brief summary of errors in force for the Adaptive TreePM code
with the new parameter $n_n$.  
Given that this code is based on the TreePM code, it inherits the errors
in force with respect to all the other parameters.  
As we have seen in earlier work 
\citep{2002JApA...23..185B,2003NewA....8..665B,2008arXiv0802.3215K}, errors in
force can be minimised to a fairly low level for the TreePM with a judicious
choice of parameters. 

We find that the time taken by the Adaptive TreePM (for $n_n=32$) is
comparable to that taken by the fixed resolution TreePM code if the latter
uses a force softening length that is about $1/20$ of the mean inter-particle
separation. 

One of the key reasons for considering adaptive resolution is to avoid two
body collisions.  
Cosmological simulations require collisionless evolution. 
We test the Adaptive TreePM by simulating collapse of an oblique plane-wave.  
We find that unlike the fixed resolution TreePM code that leads to large
transverse motions, a clear sign of two body collisions, the Adaptive TreePM
code does not have any significant transverse motions. 
(See Figures~\ref{fig_plwave_treepm}, \ref{fig_plwave_atreepm})

We have compared the performance of the Adaptive TreePM with that of the fixed
resolution TreePM code for power law initial conditions. 
We used a power law model and the Einstein-de Sitter background as it allows
us to test codes by requiring self-similar evolution of the correlation
function. 
We find that for $n_n=32$, the resulting density distribution in the Adaptive
TreePM is comparable to that seen in the highest resolution TreePM in many
respects. 
The correlation function for the two match at all scales and the mass function
of collapsed haloes matches for haloes with more than $n_n$ particles. 
The Adaptive TreePM with the $\gradeps$ term in the equation of motion does
much better than the TreePM as well as the Adaptive TreePM without the extra
term in resolving highly over-dense cores of massive haloes. 
It is noteworthy that the Adaptive code takes less time than the TreePM it has
been compared with (see fig.~\ref{fig_timing}). 

We have tested the codes by looking for convergence in results as we modify
the key parameters that describe the force resolution.  
We find that the fixed resolution TreePM code converges slower than expected
from, for example, self-similar evolution of clustering. 
The Adaptive TreePM code with the $\gradeps$ term converges fairly quickly and
offers an effective dynamical range that is slightly smaller than the fixed
resolution TreePM that takes almost the same time to run. 

Given our analysis of errors, and the comparison of the performance of the
ATreePM code with and without the extra term in the equation of motion, the
natural conclusion is that we require the extra term in order to obtain low
errors and numerical convergence. 
This raises the obvious question about the AMR codes where no such extra term
is used. 
Further, in most AMR codes, resolution is increased when there are of order
$10$ particles in the lower resolution elements. 
We find that the errors are minimized when this number is around $20$ even
when the extra term is not taken into account.  
It is not clear how serious these issues are, given that AMR codes have been
developed and tested in a variety of ways over the last three decades, but it
is a concern\footnote{One aspect where AMR codes do relatively poorly is in
  getting the halo mass function at the low mass end
  \citep{2005ApJS..160....1O,2008CS&D....1a5003H}.  
As one can see in figures in the papers cited here, the shortfall is often
spread over a decade in mass of haloes.  
We do not see any gradual decline in the number density of haloes in ATreePM
up to the cutoff set by $n_n$.}. 

We have established in this paper that an adaptive resolution code for
evolution of perturbations in collisionless dark matter can give reliable
results for a range of indicators from clustering properties to mass function
of collapsed haloes and even get the internal dynamics of collapsed haloes
tight. 
Further, we show that such a code is efficient in that it is faster than fixed
resolution codes that give us comparable force resolution in highly over dense
regions.
Such a code is very useful as it allows us to probe clustering at small scales
in a reliable manner.
Studies of box-size effects have shown that large simulation boxes are
required in order to limit the effect of perturbations at larger scales that
are not taken into account \citep{2006MNRAS.370..993B,2008arXiv0804.1197B}.  
In such a situation an adaptive code provides us with a reasonable range of
scales over which the results can be trusted.  
We expect to use this code to address several issues related to gravitational
clustering in an expanding universe. 
We also plan to revisit issues related to density profiles of collapsed
haloes. 

Given the conclusions listed above, we feel that the Adaptive TreePM methods
represents an exciting development where we can set aside worries about the
impact of collisionality. 
The relative speed of the adaptive code also makes this a more pragmatic
option.

\section*{Acknowledgments}

The authors are indebted to Daniel J. Price for clarifying several issues
regarding their implementation of the adaptive code. 
The authors thank Martin White, Volker Springel and Alessandro Romeo for
useful discussions and comments. 
Numerical experiments for this study were carried out at cluster computing
facility in the Harish-Chandra Research Institute
(http://cluster.hri.res.in).  
This research has made use of NASA's Astrophysics Data System.

\appendix
\begin{onecolumn}
\section{Expressions for Cubic Spline Softened Potential and Force Kernels}
One can integrate the Poisson equation for a given kernel to obtain 
the softened two-body potential kernel. 
For the case of the cubic spline kernel we have the expression for the 
two-body potential kernel 
$\phi({\mathbf{u}},\epsilon)$ and the regular two-body force 
${\mathbf{f}}({\mathbf{u}},\epsilon) = -\nabla_{{\mathbf{u}}}\phi$
:
\baq
  \phi({\mathbf{u}},\epsilon) &=& \left\{
  \begin{array}{ll}
    \frac{16}{\epsilon}
    \left[
      \frac{1}{3}(\frac{u}{\epsilon})^{^2} -
      \frac{3}{5}(\frac{u}{\epsilon})^{^4}
      +\frac{2}{5}(\frac{u}{\epsilon})^{^5}
      \right]-\frac{14}{5 \epsilon} ,  
    & \,\mbox{$\,\,0\,\,\leq \frac{u}{\epsilon} < 0.5$} \\
    \frac{1}{\epsilon}
    \left[
      \frac{1}{15}(\frac{\epsilon}{u})+\frac{32}{3}(\frac{u}{\epsilon})^{^2}
      -\frac{16}{1}(\frac{u}{\epsilon})^{^3}+
      \frac{48}{5}(\frac{u}{\epsilon})^{^4}
      -\frac{32}{15}(\frac{u}{\epsilon})^{^5}
      \right] -\frac{16}{5 \epsilon},
    & \,\mbox{$0.5 \leq \frac{u}{\epsilon} < 1.0$}\\
    -\frac{1}{u}, 
    & \,\mbox{$1.0 \leq \frac{u}{\epsilon}$}
  \end{array}
  \right. \label{pot_fixedh} \\
\nonumber \\
   {\mathbf{f}}({\mathbf{u}},\epsilon) &=& \left\{
  \begin{array}{ll} 
    -\frac{32{\mathbf{u}}}{\epsilon^3}
    \left[
      \frac{1}{3} -\frac{6}{5}(\frac{u}{\epsilon})^{^2}
      +(\frac{u}{\epsilon})^{^3}
      \right],  
    & \quad\qquad\mbox{$\,\,\,0\,\,\leq \frac{u}{\epsilon} < 0.5$} \\
    -\frac{{\mathbf{u}}}{\epsilon^3}
    \left[
      -\frac{1}{15}(\frac{\epsilon}{u})^{^3}+
      \frac{64}{3} - \frac{48u}{\epsilon}
      +\frac{192}{5}(\frac{u}{\epsilon})^{^2}
      -\frac{32}{3}(\frac{u}{\epsilon})^{^3}
      \right],
    & \quad\qquad\mbox{$0.5 \leq \frac{u}{\epsilon} < 1.0$} \\
    -\frac{{\mathbf{u}}}{u^3} 
    & \quad\qquad\mbox{$1.0 \leq \frac{u}{\epsilon}$}
  \end{array}
  \right. \label{force_fixedh}
\eaq
For variable or local smoothing length 
$\epsilon_i \equiv \epsilon(\mathbf{r}_i)$ there will be an additional
term given in eqs.(10-12) for which one needs to compute 
$({\partial W}/{\partial \epsilon})$,  $({\partial W}/{\partial \mathbf{u}})$
and $({\partial \phi}/{\partial \epsilon})$
For the cubic spline kernel these expressions are:
\baq
\frac{\partial W}{\partial \epsilon} &=&       
\frac{8}{\pi \epsilon^4} 
\left\{
  \begin{array}{ll}
      3\left[
	- 1  
	+ 10\left(\frac{u}{\epsilon}\right)^2 
	- 12\left(\frac{u}{\epsilon}\right)^3\right],  
      & \,\mbox{$\,\,0\,\,\leq \frac{u}{\epsilon} < 0.5$} \\
      6\left[ 
	- 1
	+ 4\left(\frac{u}{\epsilon}\right)
	- 5\left(\frac{u}{\epsilon}\right)^2
	+ 2\left(\frac{u}{\epsilon}\right)^3\right], 
      & \,\mbox{$0.5 \leq \frac{u}{\epsilon} < 1.0$}\\
      0, 
      & \,\mbox{$1.0 \leq \frac{u}{\epsilon}$}
  \end{array}
  \right. \label{kernel_dwde}\\
  \frac{\partial W}{\partial \mathbf{u}} &=& 
  \frac{48}{\pi\epsilon^4}\frac{\mathbf{u}}{u}
  \left\{
  \begin{array}{ll}
       - 2\left(\frac{u}{\epsilon}\right)
       + 3\left(\frac{u}{\epsilon}\right)^2
    & \,\mbox{$\,\,0\,\,\leq \frac{u}{\epsilon} < 0.5$} \\
      -1 
      + 2\left(\frac{u}{\epsilon}\right)
      - \left(\frac{u}{\epsilon}\right)^2
    & \,\mbox{$0.5 \leq \frac{u}{\epsilon} < 1.0$}\\
    \,\,\,\, 0,
    & \,\mbox{$1.0 \leq \frac{u}{\epsilon}$}
  \end{array}
  \right. \label{kernel_dwdr}\\
  \frac{\partial \phi}{\partial \epsilon} &=&
  \left\{   
  \begin{array}{ll}
    \frac{16}{\epsilon^2}
    \left[
      - \left(\frac{u}{\epsilon}\right)^2
      + 3\left(\frac{u}{\epsilon}\right)^4
      - \frac{12}{5}\left(\frac{u}{\epsilon}\right)^5
    \right]
    + \frac{14}{5\epsilon^2}
    & \,\mbox{$\,\,0\,\,\leq \frac{u}{\epsilon} < 0.5$} \\
    \frac{1}{\epsilon^2}
    \left[
      - 32\left(\frac{u}{\epsilon}\right)^2
      + 64\left(\frac{u}{\epsilon}\right)^3
      - 48\left(\frac{u}{\epsilon}\right)^4
      + \frac{64}{5}\left(\frac{u}{\epsilon}\right)^5
      \right]
    +\frac{16}{5\epsilon^2}
    & \,\mbox{$0.5 \leq \frac{u}{\epsilon} < 1.0$}\\
      \,\,\,\, 0,
  & \,\mbox{$1.0 \leq \frac{u}{\epsilon}$}
  \end{array}
  \right. \label{pot_dphide}
\eaq

\end{onecolumn}

\label{lastpage}

\end{document}